\pdfoutput=1
\documentclass[preprint,review,12pt,3p]{elsarticle}
\usepackage{amssymb}
\usepackage{graphicx,subfigure}
\usepackage{comment}
\usepackage{ctable}
\usepackage{caption}
\usepackage{color}

\usepackage{amssymb}
\usepackage{amsmath}
\usepackage{bbm}
\usepackage{wrapfig}
\usepackage{cases}
\usepackage{mathtools}

\usepackage[english]{babel}
\usepackage{amsmath,amsthm}
\usepackage{amsfonts}
 \usepackage{epsfig,amsthm}
 \usepackage{float}
 \usepackage{epstopdf}

\theoremstyle{definition}

\theoremstyle{remark}

\numberwithin{equation}{section}

\newcommand{\D}{\mathrm{d}}

\begin{document}

\title{Phase field model for self-climb of prismatic dislocation loops by vacancy pipe diffusion}
\author[xiamen]{Xiaohua Niu\corref{cor1}}
\cortext[cor1]{Corresponding author}
 \ead{xhniu@xmut.edu.cn}
 \author[hkust]{Yang Xiang}
 \author[uconn]{Xiaodong Yan}

\address[xiamen]{School of Applied Mathematics,  Xiamen University of Technology,  Xiamen,   361024,   China. }
\address[hkust]{Department of Mathematics, The Hong Kong University of Science and Technology, Clear Water Bay, Kowloon, Hong Kong}
\address[uconn]{Department of Mathematics, University of Connecticut, Storrs, CT 06269, United States}

\begin{abstract}

 In this paper, we present a  phase field model for the self-climb motion of prismatic dislocation loops via vacancy pipe diffusion driven by elastic interactions. This conserved dynamics model is developed under the  framework of the Cahn-Hilliard equation with incorporation of the climb force on dislocations, and is based on the  dislocation self-climb velocity formulation established in Ref.~\cite{Niu2017}. The phase field model has the advantage of being able to handle the topological and geometrical changes automatically during the simulations. Asymptotic analysis shows  that the proposed phase field model gives the dislocation self-climb velocity accurately in the sharp interface limit. Numerical simulations of evolution, translation, coalescence and repelling of prismatic loops  by self-climb show excellent agreement with   discrete dislocation dynamics simulation results and the  experimental observation.

%

\end{abstract}


%


\maketitle

\section{Introduction}


The self-climb of dislocations (line defects in crystalline materials) is driven by pipe diffusion of vacancies along the dislocations~\cite{Hirth-Lothe}.
During the self-climb motion of a prismatic dislocation loop, the enclosed area is conserved, thus such a motion is also called conservative climb of dislocation loops. The self-climb motion is the dominant
mechanism of prismatic loop motion and coalescence at not very high temperatures. Dislocation self-climb motion plays important roles in the properties of irradiated materials and has been an active research topic ever since it was first observed in 1960's \cite{Kroupa1961,Silcox1960,Washburn1960,Johnson1960,Turnbull1970,Narayan1972,DudarevReview,
DudarevLangevin,Swinburne2016,Okita2016,Niu2017,Niu2019,LiuJMPS2020,LiuMSMSE2020}.

    In Ref.~\cite{Niu2017}, a dislocation self-climb velocity formulation was obtained through upscaling from an atomistic model incorporating vacancy pipe diffusion along the dislocations. Unlike early dislocation self-climb models which gave the velocity of small circular prismatic loops, the formulation in Ref.~\cite{Niu2017} gives the self-climb velocity at every point of dislocations that could have any shape (not necessarily circular loops), and is able to maintain the area enclosed by a prismatic loop during its self-climb motion. In Ref.~\cite{Niu2019}, a numerical implementation method of this self-climb formulation for discrete dislocation dynamics (DDD)  was proposed, and simulations were performed
on evolution, translation, coalescence and other self-climb motions of prismatic loops that are in excellent agreement with experimental and atomistic results. It has also been shown that this
self-climb formulation is able to quantitatively explain the annihilation of prismatic loops with free surfaces by dislocation self-climb observed in molecular dynamics simulations \cite{Mordehai2017}  and self-healing of low angle grain boundaries by vacancy diffusion \cite{Gu2018}. An alternative derivation of the self-climb velocity based on surface diffusion and finite element framework was presented \cite{LiuJMPS2020} and improved finite element formulation was developed \cite{LiuMSMSE2020}.

The numerical method for the self-climb DDD simulations proposed in Ref.~\cite{Niu2019}   is based on the front-tracking discretization, in which dislocations are discretized into small segments and each segment is tracked during the evolution.
However, as other front-tracking methods, this DDD discretization method for self-climb is not robust to handle topological changes. For example, during the  coalescence of two prismatic loops, extremely small time step is needed to avoid numerical instability when two segments from different loops are very close to each other, and at some moment, we need to make a decision to connect these two segments manually to complete the coalescence process. Moreover, frequent re-meshing is needed during the simulations.
 For example, during the translation of a prismatic loop under a stress gradient, the numerical nodal points are moving in the opposite direction with respect to the loop motion direction and tend to be accumulated in the back side of the loop.
  In this paper, we develop a simulation method based on the phase field model to address these limitations.


The phase field method  is a commonly used framework for evolution of interfaces  \cite{LongQing2002,WangLi2010,DuFeng2020}, in which the interfaces are described in diffuse form by an order parameter function $\phi$ defined over the entire domain, and evolution of the interfaces is implicitly determined by the evolution of the function $\phi$. Phase field models are able to handle topological changes of the interfaces automatically, and can be simply implemented on a uniform mesh of the simulation domain.
   The classical Cahn-Hilliard equation with degenerate mobility was employed to model the interface motion by surface diffusion~\cite{CahnTaylor1994}, and a proof by formal asymptotic analysis was given in Ref.~\cite{Cahn1996}.
   Phase field models for the instability and evolution of solid-liquid interfaces under stress  were proposed~\cite{Muller1999,Kassner_1999,Kassner2001}, and asymptotic analysis of the sharp interface limit of the surface velocity was performed~\cite{Kassner2001}.
   In Ref.~\cite{RATZ2006187}, a phase field model was proposed for heteroepitaxial growth of thin films. The model incorporated elastic energy into  the framework of Cahn-Hilliard equation with degenerate mobility, and a stabilizing function was introduced to maintain the equilibrium values of the phase-field function $\phi$ and to enforce the correct sharp interface limit of the dynamics. Their phase field model was implemented by finite element methods. How to generate correct sharp interface limits of surface diffusion from phase field models under the framework of Cahn-Hilliard equation with degenerate mobility was discussed in Ref.~\cite{Gugenberger2008Kassner}, and  globally and locally conservative tensorial models were proposed. A phase field approach to simulating solid-state dewetting problems was developed in Ref.~\cite{JIANG20125578}. Dependence of the interfacial dynamics associated with the Cahn-Hilliard equation  on the specific forms of the nonlinear potential and the nonlinear and degenerate diffusion mobility  were examined using asymptotic analysis in Ref.~\cite{QiangDu2014}.
      In Ref.~\cite{Lee2016}, the sharp-interface limits for the degenerate Cahn –Hilliard equation were revisited  using multiple inner layers in the asymptotic analysis and two subcases were identified.

 %

       There is no phase field model available in the literature for the self-climb motion of dislocations, although a number of phase field models have been proposed for other motions of dislocations including glide and climb by vacancy bulk diffusion.  Glide of dislocations is the motion in their slip planes and climb is the motion out of the slip planes.
       In the climb by vacancy bulk diffusion, the prismatic loops shrinks or grows under climb force \cite{Hirth-Lothe,Ghoniem2000, Xiang2003,Mordehai2008,Gu2015},  which is completely different from the self-climb motion by vacancy pipe diffusion along the dislocations with which the enclosed area of a prismatic loop is conserved. The available phase field models for these dislocation motions of glide and climb by vacancy bulk diffusion take the form of the nonconservative, Allen-Cahn form instead of the conservative, Cahn-Hilliard form. Several phase field models for the glide motion of dislocations have been developed~\cite{PFdislocation2001,Rodney2003,ShenWang2003}. There are also phase field models for the climb by vacancy bulk diffusion of the prismatic loops in their climb planes  coupled with vacancy/interstitial diffusion equations \cite{Shenyang2011,Geslin2014,Ke-Yunzhi2014}.
        In \cite{Pengchuang-Sanqiang2019}, a phase field model of dislocation climb by vacancy bulk diffusion for describing the structure of low-angle symmetrical tilt grain boundaries was developed and grain boundaries sink strength was evaluated.

      In this paper, we present a new phase field model  for the self-climb of prismatic dislocation loops by vacancy pipe diffusion based on the  dislocation self-climb velocity formulation obtained in \cite{Niu2017}.  We focus on the two dimensional problem where the prismatic dislocation loops  lie and evolve by self-climb in a climb plane.  This phase field model is based on the framework of the Cahn-Hilliard equation with incorporation of the climb force on dislocations.  The phase field model has the advantage to handle topological  automatically and to avoid re-meshing during simulations. We perform asymptotic analysis to show  that our phase field model yields the correct self-climb velocity  in the sharp interface limit.  In particular, we show that incorporation of the climb force in the framework of the classical Cahn-Hilliard equation  does not lead to significant change in the interface profile  (dislocation core profile here) of  the classical Cahn-Hilliard phase field model, and the framework of the phase field model gives an extra term that serves to correct the dislocation climb force calculated by the phase field formulation. We validate our phase field model by numerical simulations and comparisons with the DDD simulation results obtained in  Ref.~\cite{Niu2019} and available  experimental observations.

      The rest of the paper is organized as follows. In Sec.~\ref{sec:velocity}, we review the dislocation self-climb formulation obtained in Refs.~\cite{Niu2017, Niu2019}. In Sec.~\ref{sec:pf}, we present our phase field model for self-climb of prismatic dislocations in a climb plane. In Sec.~\ref{sec:asymptotic}, we perform asymptotic analysis to show the proposed phase field model is able to give the dislocation self-climb velocity formulation in the sharp interface limit. In Sec.~\ref{sec:simulation}, we perform numerical simulations using our phase field model for the  evolution, translation and  coalescence of prismatic loops by self-climb and compare the results with  those of DDD simulation and experiments. Conclusions and discussion are made in Sec.~\ref{sec:conclusions}.

 \section{Velocity formulation of dislocation self-climb by vacancy pipe diffusion}\label{sec:velocity}
In this section, we briefly review the velocity formulation of dislocation self-climb by vacancy pipe diffusion established in Refs.~\cite{Niu2017,Niu2019}.



At room temperature,  the self-climb due to vacancy pipe diffusion along the dislocations is dominant in the dislocation climb process.  When the climb force is not very large, the self-climb velocity is~\cite{Niu2017,Niu2019}
\begin{equation}\label{eqn.self-climb-v}
v_{\rm cl}^{\rm (p)}=-\frac{c_0D_c \Omega}{k_BT}\frac{\D^2 f_{\rm cl}}{\D s^2},
\end{equation}
where $D_c$ is the vacancy diffusion constant in the dislocation core region,   $s$ is the arclength parameter along the dislocation line, $c_0$ is the reference equilibrium vacancy concentration on the dislocation, $k_B$ is Boltzmann's constant, $T$ is the  temperature, $\Omega$ is the volume of an atom, and $f_{\rm cl}$ is the climb component of the Peach-Koehler force on the dislocation. From Eq.~\eqref{eqn.self-climb-v}, $\int_\gamma v_{\rm cl} ds= 0$ for any dislocation loop $\gamma$. This means that the area enclosed by any dislocation loop is conserved during the self-climb motion~\cite{Niu2017}, which is an important feature of the self-climb motion of dislocations~\cite{Kroupa1961,Hirth-Lothe}. (The self-climb motion is also called conservative climb in this sense.)

The climb  Peach-Koehler force $f_{\rm cl}$  is given by
 \begin{equation}\label{eqn:climbforce00}
f_{\rm cl}=\mathbf{f}_{\rm PK}\cdot \mathbf l_{\rm cl},
\end{equation}
where
$\mathbf{f}_{\rm PK}$ is the Peach-Koehler force on dislocation calculated by~\cite{Hirth-Lothe}
\begin{equation}
\mathbf{f}_{\rm PK}=(\boldsymbol{\sigma}\cdot \mathbf{b})\times \boldsymbol{\xi},
\end{equation}
 $\mathbf b$ is the Burgers vector of the dislocation, $\boldsymbol{\sigma}$ is the stress tensor, $\boldsymbol{\xi}$ is the dislocation line direction, and
$\mathbf l_{\rm cl}$ is the climb direction:
\begin{equation}\label{eqn:climbdirection}
\mathbf l_{\rm cl}=\boldsymbol{\xi}\times\mathbf{b}/b,
\end{equation}
which is the normal direction of the slip plane of the dislocation, i.e., the plane that contains the dislocation line direction $\boldsymbol{\xi}$ and the Burgers vector $\mathbf{b}$. Here $b=|\mathbf{b}|$ is the length of the Burgers vector.
Accordingly, we have the self-climb velocity vector:
\begin{equation}\label{climbdirections}
\mathbf v_{\rm cl}^{\rm (p)}=v_{\rm cl}^{\rm (p)}\mathbf l_{\rm cl}.
\end{equation}

The stress $\boldsymbol{\sigma}$ includes the stress generated by all the dislocations and the stress from other origins, e.g. the applied stress. The stress field in general can be obtained by solving the elasticity problem in the material~\cite{Hirth-Lothe,Landau}.

In this paper, we focus on the two dimensional problem, in which prismatic dislocation loops  lie and evolve by self-climb in the $xy$ plane. Note that a prismatic loop is a dislocation loop whose  Burgers vector  is  perpendicular to the plane that contains the loop. In this two dimensional setting, all the prismatic loops have the same Burgers vector in  the $z$ direction:  $\boldsymbol{\rm b}=(0, 0, b)$, and the directions of their climb motion and the climb force on them are both in the $xy$ plane. In this case, the climb Peach-Koehler force on dislocations given by Eqs.~\eqref{eqn:climbforce00} and \eqref{eqn:climbdirection} can be written as
\begin{eqnarray}\label{eqn:climbforce11}
f_{\rm cl}=-\sigma_{33}b.
 \end{eqnarray}
 Here $f_{\rm{cl}}$  is the total climb force
 \begin{equation}\label{eqn:totalforce}
 f_{\rm{cl}}=f_{\rm{cl}}^{\rm d}+f_{\rm{cl}}^{\rm app},
 \end{equation}
 where $f_{\rm{cl}}^{\rm d}=-\sigma_{33}^{\rm d}b$ is the climb force generated by all the dislocations with $\sigma_{33}^{\rm d}$ being the stress component generated by all the dislocations,
and $f_{\rm{cl}}^{\rm app}=-\sigma_{33}^{\rm app}b$ with $\sigma_{33}^{\rm app}$ being a component of the applied stress.
When the stress field is only generated by all the dislocation loops in the $xy$ plane, at a point $\left(x, y\right)$ on the plane, the stress component $ \sigma_{33}$ is~\cite{Hirth-Lothe}
\begin{eqnarray}\label{sigma33}
\sigma_{33}^{\rm d}\left(x, y\right)=\frac{\mu b}{4\pi\left(1-\nu\right)}\int_{C}\frac{y-\bar{y}}{R^3}d\bar{x}-\frac{x-\bar{x}}{R^3}d\bar{y},
\end{eqnarray}
where $\mu$ is the shear modulus, $\nu$ is the Poisson ratio, $C$ is the collection of all the dislocation loops, the point $(\bar{x}, \bar{y})$ varies along $C$, and $R=\sqrt{\left(x-\bar{x}\right)^2+\left(y-\bar{y}\right)^2}$.

It can be seen that climb force $f_{\rm cl}$ given by Eqs.~\eqref{eqn:climbforce11} and \eqref{sigma33} is long-range (nonlocal), meaning that it depends on all the points on   these dislocation loops.
     When the point $(x,y)$ is approaching the dislocation $C$, the following asymptotic behavior holds~\cite{GB,ZhaoXiang}
\begin{eqnarray}\label{sigma33-appro}
\sigma_{33}^{\rm d}\left(x, y\right)=\dfrac {\mu b  }{2\pi \left( 1-\nu \right) d}-\dfrac {\mu b}{4\pi \left( 1-\nu \right) }\kappa \ln |\rho|+O\left( 1\right),
\end{eqnarray}
where $\kappa$ is the curvature of the dislocation in an anticlockwise direction, $d$ is the signed distance from point $(x,y)$ to the dislocation $C$, with ``$+$" sign when the point $(x, y)$ is on the $\mathbf b\times \boldsymbol{\xi}$-side of $C$ and ``$-$" sign on the other side. Note that direct calculation using the stress formula in Eq.~\eqref{sigma33} will lead to singularity on the dislocations; see also Eq.~\eqref{sigma33-appro} as $d\rightarrow 0$.  There are several methods to avoid this non-physical singularity caused by linear elasticity theory, and one  method is to spread the
Burgers vector  over the core region of the dislocations incorporating the nonlinear effect in the dislocation core~\cite{Hirth-Lothe,Lothe1992,Xiang2003,Cai2006,ZhaoXiang}. This treatment will be adopted in this paper. Any of such treatments will give the following leading order behavior of the Peach-Koehler force on the dislocation~\cite{GB}:
\begin{eqnarray}\label{force-appro}
f_{\rm cl}^{\rm d}=\dfrac {\mu b^2}{4\pi \left( 1-\nu \right) }\kappa \ln r_d+O\left( 1\right),
\end{eqnarray}
where $r_d$ is the core radius of the dislocations and is small compare with the size of the dislocation loop (or the radius of curvature at that point on the dislocation). Recall that $f_{\rm cl}^{\rm d}$ is in the direction of $\mathbf l_{\rm cl}$ defined in Eq.~\eqref{eqn:climbdirection}.

Moreover, using a regularized Dirac delta function $\delta_C(x,y)$ of the dislocation $C$ ( i.e., at each point on $C$, $\delta_C(x,y)$ is a regularized one dimensional Dirac delta function in the cross-section perpendicular to $C$) to represent the spreading of the Burgers vector   over the dislocation core region as described above, the stress component $ \sigma_{33}$ in Eq.~\eqref{sigma33} can be written as
\begin{eqnarray}\label{sigma33_reg}
\sigma_{33}^{\rm d}\left(x, y\right)=\frac{\mu b}{4\pi\left(1-\nu\right)}\int_{\Omega}\left(\frac{y-\bar{y}}{R^3}\xi_1-\frac{x-\bar{x}}{R^3}\xi_2\right)
\delta_C(\bar{x},\bar{y})d\bar{x}d\bar{y},
\end{eqnarray}
where the dislocation line direction $\boldsymbol{\xi}=(\xi_1,\xi_2)$. The width of the regularized Dirac delta function $\delta_C(x,y)$ is the dislocation core width $2r_d$. Using the regularized Dirac delta function $\delta_C(x,y)$, the asymptotic behavior of the stress in Eq.~\eqref{sigma33-appro} becomes
\begin{eqnarray}\label{eqn.elasiticity}
\sigma_{33}^{\rm d}(d)=\dfrac {\mu b  }{2\pi \left( 1-\nu \right)}\int^{+\infty}_{-\infty}\frac{\delta_C(\eta)}{d-\eta}{\rm d}\eta-\dfrac {\mu b}{4\pi \left( 1-\nu \right) }\kappa \ln r_d+O\left( 1\right), \ \ d\rightarrow 0,
\end{eqnarray}
where $d$ and $\eta$ are the signed distance to  the dislocation in the cross-section perpendicular to the dislocation, with ``$+$" sign when the point is on the $\mathbf b\times \boldsymbol{\xi}$-side of the dislocation.

%

%

In Ref.~\cite{Niu2019}, numerical method for discrete dislocation dynamics (DDD) based on the self-climb formulation in Eq.~\eqref{eqn.self-climb-v} has been developed using front-tracking discretization,   and simulations have been performed using this numerical method on self-climb of dislocations and the results agree quantitatively  with the available experimental observations and atomistic simulations, including evolution, translation and coalescence of prismatic loops as well as prismatic loops driven by an edge dislocation by self-climb motion and the elastic interaction between them.   In this DDD discretization method, dislocations are discretized into small segments and each segment is tracked during the evolution.
However, as other front-tracking methods, this DDD discretization method for self-climb is not robust to handle topological changes, and  frequent re-meshing is needed during the simulations; see the examples described in the introduction.

 In this paper,  we will develop a simulation method based on the phase field model for the dislocation self-climb motion. The developed phase field model will address these limitations of the front-tracking based method, and especially, topological changes will be handled automatically and there is no need for re-meshing. Formulation of the phase field model will be presented in the next section.

\section{Phase field model for dislocation self-climb}\label{sec:pf}

 In this section, we present our phase field model for the self-climb motion of dislocations in a plane. The phase field model is based on the classical Cahn-Hilliard equation with degenerate mobility~\cite{CahnTaylor1994,Cahn1996} and incorporates
  the dislocation self-climb velocity in Eq.~\eqref{eqn.self-climb-v}. Recall that the Cahn-Hilliard equation with degenerate mobility describes the process of phase separation by surface diffusion, and the velocity of the interface in the sharp interface limit in the two dimensional setting is proportional to $-\frac{\D^2 \kappa}{\D s^2}$, where $\kappa$ is the curvature of the interface and $s$ is its arclength parameter. This sharp interface limit velocity is local, and is quite different from the dislocation self-climb velocity in Eq.~\eqref{eqn.self-climb-v} with the nonlocal climb force $f_{\rm cl}$ in Eqs.~\eqref{eqn:climbforce11}
and \eqref{sigma33}. By using the asymptotic behavior of the climb force in Eq.~\eqref{force-appro}, we will show that an appropriate linear combination of these two driving forces is able to give an accurate dislocation self-climb velocity in Eq.~\eqref{eqn.self-climb-v} when the self-climb force $f_{\rm cl}$  is incorporated into the Cahn-Hilliard equation with degenerate mobility.

In the phase field model, we use an order parameter $\phi$ defined in the entire domain $\Omega$  to describe the dislocations, which are the interfaces between the regions of two stable values of $\phi=0$ and $\phi=1$. We first focus on the vacancy loops, for which $\phi=0$ inside the loops and $\phi=1$ outside.
  The local dislocation line direction is defined as $\boldsymbol{\xi}=\frac{\mathbf b}{b}\times\frac{\nabla \phi}{|\nabla \phi|}=(-\phi_y,\phi_x)\frac{1}{|\nabla \phi|}$ in the $xy$ plane, and these dislocation loops are in the counterclockwise direction with resect to the $z$ axis. In fact, $|\nabla \phi|$ can be considered as a regularized Dirac delta function in the cross-section at a point on a dislocation, and we have $(-\phi_y,\phi_x)=\boldsymbol{\xi}|\nabla \phi|$.
In the phase field model, using this property of $\phi$ and following Eqs.~\eqref{eqn:climbforce11} and \eqref{sigma33_reg},  we can write the climb force $f_{\rm{cl}}$ generated by all the dislocations  as
\begin{equation}\label{eqn:fcl_pf}
f_{\rm{cl}}^{\rm d}\left(x, y,\phi\right)=\frac{\mu b^2}{4\pi\left(1-\nu\right)}\int_\Omega\left(\frac{x-\bar{x}}{R^3}
\phi_{\bar{x}}+\frac{y-\bar{y}}{R^3}\phi_{\bar{y}}\right)d\bar{x}d\bar{y}.
\end{equation}
Note that this $f_{\rm{cl}}\left(x, y\right)$ is defined everywhere in the domain $\Omega$.

The phase field model for the dislocation self-climb motion is
 \begin{flalign}\label{eqn.phasefield}
     \phi_t =&  \nabla \cdot \left(M \left(\phi\right)  \nabla \frac{\mu_c }{ g\left(\phi\right) }\right), \\
     \mu_c =& -\nabla^2\phi+\frac{1}{\varepsilon^2}q'\left(\phi\right)+\frac{1}{\varepsilon} h\left(\phi\right)f_{\rm{cl}},\label{eqn:mu_c}
    \end{flalign}
    where $\varepsilon$ is a small parameter that controls the width of the interface between the regions $\phi=0$  and $\phi=1$. This means that  the core diameter of the dislocations is of $O(\varepsilon)$.

 In this model, $\mu_c$ is the chemical potential, which comes from variations of the total energy that includes the energy in the classical Cahn-Hilliard equation and the elastic energy due to dislocations. The first two terms in $\mu_c$ in Eq.~\eqref{eqn:mu_c} is the variation of the energy in the classical Cahn-Hilliard equation $E_{\rm CH}(\phi)$: $-\nabla^2\phi+\frac{1}{\varepsilon^2}q'\left(\phi\right)=\frac{\delta E_{\rm CH}}{\delta \phi}$, where
 \begin{equation}
 E_{\rm CH}(\phi)=\int_{\Omega}\left(\frac{1}{2} |\nabla \phi |^2+\frac{1}{\varepsilon^2}q(\phi)\right)d\Omega,
  \end{equation}
  and $q(\phi)$ is a double-well potential that has the two energy minimum state $\phi=0$ and $\phi=1$. As other phase field models, we choose $q(\phi)$ to be
  \begin{equation}
  q\left(\phi\right)=2\phi^2\left(1-\phi\right)^2.
   \end{equation}
Here with the factor $2$ in this expression, it can be calculated that for a straight dislocation, the dislocation core radius is approximately $\varepsilon$ without the term of climb force $f_{\rm{cl}}$, and with the $f_{\rm{cl}}$ term, is about $1.75\varepsilon$.

In Eq.~\eqref{eqn:mu_c}, the climb force $f_{\rm{cl}}$  is added in the chemical potential $\mu_c$ of the Cahn-Hilliard equation at the order of $1/\varepsilon$. As given in Eq.~\eqref{eqn:totalforce}, the total climb force $f_{\rm{cl}}$ includes  $f_{\rm{cl}}^{\rm d}$, which is the climb force generated by all the dislocations as given in Eq.~\eqref{eqn:fcl_pf}, and the climb force due to the applied stress $f_{\rm{cl}}^{\rm app}=-\sigma_{33}^{\rm app}b$.
  The purpose of the coefficient function $h(\phi)$ of the climb force $f_{\rm{cl}}$   is to add $f_{\rm{cl}}$ only on the dislocations, i.e., $h(\phi)=0$ when $\phi=1$ and $0$.   We choose
  $h(\phi)=\frac{H_0}{2} q\left(\phi\right)=H_0\phi^2(1-\phi)^2$, where $H_0$ a constant factor.  The constant $H_0$ is chosen such that the phase field model generates accurate self-climb velocity of the dislocations as given in Eq.~\eqref{eqn:climbforce00};  see  Sec.~\ref{sec:inner} for details.
  The climb force $f_{\rm{cl}}$ can be written as the variation of the elastic energy, i.e. $f_{\rm{cl}}=-\frac{\delta E_{\rm el}}{\delta \phi}$, where the elastic energy
 \begin{equation}
  E_{\rm el}=\int_\Omega\left(-\frac{1}{2}\phi f_{\rm{cl}}^{\rm d}-\phi f_{\rm{cl}}^{\rm app}\right) d\Omega.
  \end{equation}
  The chemical potential $\mu_c$ in Eq.~\eqref{eqn:mu_c} can be written as
  \begin{equation}
  \mu_c=\frac{\delta E_{\rm CH}}{\delta \phi}+\frac{1}{\varepsilon}h(\phi)\frac{\delta E_{\rm el}}{\delta \phi}.
  \end{equation}

 The function $M(\phi)$ in the phase field model in Eq.~\eqref{eqn.phasefield} is the degenerate mobility~\cite{CahnTaylor1994,Cahn1996}, which vanishes when $\phi=1$ and $0$.
  The function $g(\phi)$ is the stabilizing function proposed in Ref.~\cite{RATZ2006187}, which also vanishes when $\phi=1$ and $0$.   The purpose of the   singular factor $1/g(\phi)$ of the chemical potential is to
    guarantee correct asymptotics of surface diffusion in the sharp interface limit~\cite{RATZ2006187,Gugenberger2008Kassner}. We choose
   $    g(\phi)=\frac{1}{2}q(\phi)=\phi^2(1-\phi)^2$, and $ M(\phi)=M_0g(\phi)$
     where  constant $M_0$ is chosen such that the phase field model gives the correct dislocation self-climb velocity; see Sec.~\ref{sec:inner} for details.

The integral formulation of $f_{\rm{cl}}^{\rm d}(x,y,\phi)$ in Eq.~\eqref{eqn:fcl_pf} has a simpler expression in the Fourier space under periodic boundary conditions. In this case,
the Fourier transformation of $f_{\rm{cl}}^{\rm d}(x,y,\phi)$ in terms of $e^{i\omega_1x}e^{i\omega_2y}$ is
\begin{equation}\label{eqn:fftfcl}
\hat{f_{\rm{cl}}^{\rm d}}\left(\omega_1,\omega_2,\phi\right)=\frac{\mu b^2}{2\left(1-\nu\right)}  \sqrt{\omega^2_1+\omega^2_2}\ \hat{\phi},
\end{equation}
where $\omega_1$ and $\omega_2$ are frequencies in the $x$ and $y$ directions, respectively, and $\hat{\phi}$ is the Fourier transform of $\phi$. This formulation of $f_{\rm{cl}}^{\rm d}$ in Fourier space is local and does not have singularity, thus it is more convenient to use in numerical simulations.

Recall that the phase field model for dislocation self-climb in Eqs.~\eqref{eqn.phasefield} and \eqref{eqn:mu_c} is for
 vacancy loops, with $\phi=0$ inside the loops and $\phi=1$ outside, and the loops are in the counterclockwise direction. When both vacancy loops and interstitial loops, i.e., both counterclockwise and clockwise loops, are present, we use two order parameters $\phi_1$ and $\phi_2$ for these two types of prismatic loops, respectively. For the counterclockwise (vacancy) loops, $\phi_1=0$ inside the loops and $\phi_1=1$ outside;
 for the clockwise (interstitial) loops, $\phi_2=1$ inside the loops and $\phi_2=0$ outside. The phase field model for dislocation self-climb when both types of prismatic loops are present, is
\begin{flalign}\label{eqn.phasefield_multi1}
     {\phi_1}_t =&  \nabla \cdot \left(M \left(\phi_1\right)  \nabla \frac{{\mu_c}_1 }{ g\left(\phi_1\right) }\right), \\
     {\mu_c}_1 =& -\nabla^2\phi_1+\frac{1}{\varepsilon^2}q'\left(\phi_1\right)+\frac{1}{\varepsilon} h\left(\phi_1\right)f_{\rm{cl}},\label{eqn:mu_c1}
    \end{flalign}
    and
\begin{flalign}\label{eqn.phasefield_multi2}
     {\phi_2}_t =&  \nabla \cdot\left( M \left(\phi_2\right)  \nabla \frac{{\mu_c}_2 }{ g\left(\phi_2\right) }\right), \\
     {\mu_c}_2 =& -\nabla^2\phi_2+\frac{1}{\varepsilon^2}q'\left(\phi_2\right)+\frac{1}{\varepsilon} h\left(\phi_2\right)f_{\rm{cl}},\label{eqn:mu_c2}
    \end{flalign}
 where the climb force generated by all the dislocations is
 \begin{equation}\label{eqn:fcl_pf_multi}
f_{\rm{cl}}^{\rm d}\left(x, y,\phi_1,\phi_2\right)=\frac{\mu b^2}{4\pi\left(1-\nu\right)}\int_\Omega\left[\frac{x-\bar{x}}{R^3}
\big({\phi_1}_{\bar{x}}+{\phi_2}_{\bar{x}}\big)+\frac{y-\bar{y}}{R^3}\big({\phi_1}_{\bar{y}}+{\phi_2}_{\bar{y}}
\big)\right]d\bar{x}d\bar{y}.
\end{equation}
and the total climb force $f_{\rm{cl}}$ is given by Eq.~\eqref{eqn:totalforce}.

It is tempting to use a single order parameter $\phi$ and a potential with multiple stable states, e.g. $q(\phi)=\frac{1}{\pi^2}(1-\cos(2\pi\phi))$ in the phase field self-climb model for both vacancy and interstitial loops. However, numerical examples show that such a model requires a very fine mesh to resolve the repulsive motion of two loops with different orientations. This is because when two loops with different orientations are not well-separated, the  core regions of the two loops near the closest point have stronger influence on each other than the self-climb velocities manifested at higher order in the expansion in terms of the small parameter $\varepsilon$.
 Considering the fact that the phase field model is a $4$-th order equation,  a very small time step of order $(\Delta x)^4$, where $\Delta x$ is the spatial numerical grid constant, is  required due to the CFL condition. Instead, we propose to use two order parameters  $\phi_1$ and $\phi_2$ for these two types of prismatic loops with separate evolution equations in the phase field model; see Eqs.~\eqref{eqn.phasefield_multi1}-\eqref{eqn:mu_c2}. In this model, the interaction between counterclockwise and clockwise loops is only through the long-range climb force field $f_{\rm{cl}}$ in Eqs.~\eqref{eqn:fcl_pf_multi} and \eqref{eqn:totalforce} and with a smooth cutoff factor $h(\phi_1)$ or $h(\phi_2)$, and the dislocation cores of loops with different orientations  do not interact directly with each other. See a numerical simulation using Eqs.~\eqref{eqn.phasefield_multi1}-\eqref{eqn:mu_c2}  in Sec.~\ref{sec:simulation}.

This developed phase field model has the advantage to handle topological changes automatically and to avoid re-meshing during simulations. In the next section,  we will perform asymptotic analysis to show that this phase field model gives the dislocation self-climb velocity in Eq. (\ref{eqn.self-climb-v}) in the sharp interface lime $\varepsilon\rightarrow 0$.

 \section{Asymptotic analysis of phase field model}\label{sec:asymptotic}

 We perform a formal asymptotic analysis to obtain the dislocation self-climb velocity of the proposed phase field model  in Eqs.~\eqref{eqn.phasefield} and \eqref{eqn:mu_c} in the sharp interface lime $\varepsilon\rightarrow 0$. As the asymptotic analysis for the available phase field models \cite{Cahn1996,Kassner2001,RATZ2006187,Gugenberger2008Kassner,Lee2016}, it consists of an outer expansion that describes the fields far away  from the interface and an inner expansion that represents the fields in the vicinity of the interface. The velocity of the interface is determined by asymptotic matching of these two expansions.


 \subsection{Outer expansion}

 We first perform the outer expansion in terms of the small parameter $\varepsilon$ for $\phi$ in the region away from the dislocations. Suppose that  expansion is
  \begin{eqnarray}
  \phi\left(x,y, t\right)&=&\phi^{\left(0\right)}\left(x,y, t\right)+\phi^{\left(1\right)}\left(x,y, t\right)\varepsilon+\phi^{\left(2\right)}\left(x,y, t\right)\varepsilon^2+\cdots.
  \end{eqnarray}
 Accordingly, we have
  \begin{flalign}
  g\left(\phi\right)&=g\left(\phi^{\left(0\right)}\right)+g' \left(\phi^{\left(0\right)}\right)\phi^{\left(1\right)}\varepsilon+\left(g' \left(\phi^{\left(0\right)}\right)\phi^{\left(2\right)}+\frac{1}{2}g^{''}\left(\phi^{\left(0\right)}\right)
  \left(\phi^{\left(1\right)}\right)^2\right)\varepsilon^2+\cdots,\nonumber\\
  h\left(\phi\right)&=h\left(\phi^{\left(0\right)}\right)+h'\left(\phi^{\left(0\right)}\right)\phi^{\left(1\right)}\varepsilon
  +\left(h'\left(\phi^{\left(0\right)}\right)\phi^{\left(2\right)}+\frac{1}{2}h''\left(\phi^{\left(0\right)}\right)
  \left(\phi^{\left(1\right)}\right)^2\right)\varepsilon^2+\cdots,\nonumber\\
  q'\left(\phi\right)&=q'\left(\phi^{\left(0\right)}\right)+q^{'''}\left(\phi^{\left(0\right)}\right)\phi^{\left(1\right)}\varepsilon
  +\left(q''\left(\phi^{\left(0\right)}\right)\phi^{\left(2\right)}+\frac{1}{2}q^{'''}\left(\phi^{\left(0\right)}\right)
  \left(\phi^{\left(1\right)}\right)^2\right)\varepsilon^2+\cdots,\nonumber\\
  f_{\rm{cl}}^{\rm d}(x,y,\phi)&=f_{\rm{cl}}^{\rm d}(x,y,\phi^{(0)})+ f_{\rm{cl}}^{\rm d}(x,y,\phi^{(1)})\varepsilon +f_{\rm{cl}}^{\rm d}(x,y,\phi^{(2)})\varepsilon^2+\cdots,\nonumber
   \end{flalign}
and $M(\phi)=M_0 g(\phi)$ has a similar expansion.   Note that
$f_{\rm{cl}}^{\rm d}\left(x, y,\phi\right)$ in Eq.~\eqref{eqn:fcl_pf} is linear in $\phi$. We also expand the chemical potential $\mu_c$ as
 \begin{flalign}\label{eqn:mcexpansion}
 \mu_c=\frac{1}{\varepsilon^2}\left(\mu_c^{(0)}+\mu_c^{(1)}\varepsilon+\mu_c^{(2)}\varepsilon^2+\cdots\right).
  \end{flalign}

For the convenience of expansion, we write Eq.~\eqref{eqn.phasefield} as
\begin{eqnarray}\label{eqn.phasefield1}
\phi_t=M_0\left(\Delta  \mu_c-\nabla\cdot \left( \mu_c \frac{g' \left(\phi\right)}{g\left(\phi\right)}\nabla \phi\right)\right).
\end{eqnarray}
Here we have used $M(\phi)=M_0g(\phi)$.

The $O\left(1/\varepsilon^2\right)$ equation given by Eqs.~\eqref{eqn.phasefield1} and \eqref{eqn:mu_c} is
\begin{eqnarray}\label{eq-outer-1/2order}
\Delta \mu_c^{\left(0\right)}-\nabla \cdot \left(\left( \mu_c \frac{g' \left(\phi\right)}{g\left(\phi\right)}\right)^{\left(0\right)}\nabla \phi^{\left(0\right)}\right)=0,
\end{eqnarray}
where $ \mu_c^{\left(0\right)}=q'\left(\phi^{\left(0\right)}\right)$, $ \left( \mu_c \frac{g' \left(\phi\right)}{g\left(\phi\right)}\right)^{\left(0\right)}=\frac{q'\left(\phi^{\left(0\right)}\right)g' \left(\phi^{\left(0\right)}\right)}{g\left(\phi^{\left(0\right)}\right)}=8\left(1-2\phi^{\left(0\right)}\right)^2$.
If $\phi^{\left(0\right)}=1$ or $0$, Eq.~\eqref{eq-outer-1/2order}  holds.

Next, we consider the  $O(1/\varepsilon)$ equation, which is
\begin{eqnarray}\label{eq-outer-1order}
\Delta  \mu_c^{\left(1\right)}-\nabla \cdot \left(\left(\mu_c \frac{g' \left(\phi\right)}{g\left(\phi\right)}\right)^{\left(0\right)}\nabla \phi^{\left(1\right)}\right)-\nabla\cdot\left(\left( \mu_c \frac{g' \left(\phi\right)}{g\left(\phi\right)}\right)^{\left(1\right)}\nabla \phi^{\left(0\right)}\right)=0.
\end{eqnarray}
Since  $\nabla \phi^{\left(0\right)}=0$ when $ \phi^{\left(0\right)}=0\ \text{or}\ 1$ , the last term of the above equation vanishes. Substituting $\mu_c^{\left(1\right)}=q''\left(\phi^{\left(0\right)}\right)\phi^{\left(1\right)}
+h\left(\phi^{\left(0\right)}\right)f^{\rm d}_{\rm{cl}}\left(x,y,\phi^{\left(0\right)}\right)$ into Eq.~\eqref{eq-outer-1order}, we have
\begin{eqnarray}\Delta \left(q''\left(\phi^{\left(0\right)}\right)\phi^{\left(1\right)}\right)-\nabla \cdot \left(8\left(1-2\phi^{\left(0\right)}\right)^2\nabla \phi^{\left(1\right)}\right)=0.
\end{eqnarray}
Obviously $\phi^{\left(1\right)}=0$ satisfies this equation.

The order $O(1)$ equation is
\begin{flalign}\nonumber
\phi_t^{(0)}=&M_0\left[\Delta \mu_c^{\left(2\right)}-\nabla \cdot \left(\left( \mu_c \frac{g' \left(\phi\right)}{g\left(\phi\right)}\right)^{\left(2\right)}\nabla \phi^{\left(0\right)}\right)-\nabla\cdot\left(\left( \mu_c \frac{g' \left(\phi\right)}{g\left(\phi\right)}\right)^{\left(1\right)}\nabla \phi^{\left(1\right)}\right)\right.\\
&\left.-\nabla\cdot\left(\left( \mu_c \frac{g' \left(\phi\right)}{g\left(\phi\right)}\right)^{\left(0\right)}\nabla \phi^{\left(2\right)}\right)\right].\label{eq-outer-0order}
\end{flalign}
If $\phi^{\left(0\right)}=1$ or $0$ and $\phi^{\left(1\right)}=0$,  the above equation becomes
\begin{equation}\label{eq-outer-0order-1}
\Delta  \mu_c^{\left(2\right)}-\nabla\cdot\left(\left(\mu_c \frac{g' \left(\phi\right)}{g\left(\phi\right)}\right)^{\left(0\right)}\nabla \phi^{\left(2\right)}\right)=0.
\end{equation}
 Substituting $\mu_c^{\left(2\right)}=-\Delta \phi^{\left(0\right)}+q^{''}\left(\phi^{\left(0\right)}\right)\phi^{\left(2\right)}
 +\frac{1}{2}q^{'''}(\phi^{\left(0\right)})(\phi^{\left(1\right)})^2
 +h\left(\phi^{\left(0\right)} \right)f^{\rm d}_{\rm{cl}}\left(x,y,\phi^{\left(1\right)} \right)\\
 +h'\left(\phi^{\left(0\right)} \right)\phi^{\left(1\right)}f^{\rm d}_{\rm{cl}}\left(x,y,\phi^{\left(0\right)} \right)$ and $\phi^{\left(0\right)}=1$ or $0$, $\phi^{\left(1\right)}=0$  into Eq.~\eqref{eq-outer-0order-1}, we have
\begin{eqnarray}\label{eq-outer-0order-2}\nonumber
\Delta  \left(q^{''}\left(\phi^{\left(0\right)}\right)\phi^{\left(2\right)}\right)-\nabla\cdot\left(\left( \mu_c \frac{g' \left(\phi\right)}{g\left(\phi\right)}\right)^{\left(0\right)}\nabla \phi^{\left(2\right)}\right)=0.
\end{eqnarray}
Obviously $\phi^{\left(2\right)}=0$ satisfies this equation.

In general, the equation on $O(\varepsilon^k)$, $k\geq1$, is
\begin{eqnarray}\label{eq-outer-korder}\nonumber
\phi_t^{(k)}=M_0\left[\Delta \mu_c^{\left(k+2\right)}-\sum_{i=1}^{k+2}\nabla \cdot \left(\left(\mu_c \frac{g' \left(\phi\right)}{g\left(\phi\right)}\right)^{\left(i\right)}\nabla \phi^{\left(k+2-i\right)}\right)\right].
\end{eqnarray}
If $\phi^{\left(0\right)}=1$ or $0$, and $\phi^{\left(1\right)}=\cdots=\phi^{\left(k+1\right)}=0$,  the above equation becomes
\begin{eqnarray}\label{eq-outer-korder-1}
\Delta \mu_c^{\left(k+2\right)}-\nabla\cdot\left(\left( \mu_c \frac{g' \left(\phi\right)}{g\left(\phi\right)}\right)^{\left(0\right)}\nabla \phi^{\left(k+2\right)}\right)=0.
\end{eqnarray}
Substituting $ \mu_c^{\left(k+2\right)}=-\Delta \phi^{\left(k\right)}+
\left(q^{'}\left(\phi\right)\right)^{\left(k+2\right)}
+\big(h\left(\phi\right)f^{\rm d}_{\rm{cl}}\left(x,y,\phi\right)\big)^{\left(k+1\right)}$ into Eq.~\eqref{eq-outer-korder-1}, and using
$\phi^{\left(0\right)}=1$ or $0$, $\phi^{\left(1\right)}=\cdots=\phi^{\left(k+1\right)}=0$, we have
$\left(q^{'}\left(\phi\right)\right)^{\left(k+2\right)}= q^{''}\left(\phi^{\left(0\right)}\right) \phi^{\left(k+2\right)}$
  and $f^{\rm d}_{\rm{cl}}\left(x,y,\phi^{\left(i\right)}\right)=0$ for $i=0,1, \cdots, k+1$, and
  Eq.~\eqref{eq-outer-korder-1} becomes
  \begin{eqnarray} \label{eq-outer-korder-1-1}
  \Delta q^{''}\left(\phi^{\left(0\right)}\right) \phi^{\left(k+2\right)}-\nabla\cdot\left(\left(\mu_c \frac{g' \left(\phi\right)}{g\left(\phi\right)}\right)^{\left(0\right)}\nabla \phi^{\left(k+2\right)}\right)=0.
\end{eqnarray}
 It can be seen that  $\phi^{\left(k+2\right)}=0$ is a solution.

 In summary, the outer expansion of the solution $\phi$ is
 \begin{equation}
 \phi^{\left(0\right)}=1 \  {\rm or}\ 0, \ \phi^{\left(k\right)}=0, \ k=1,2,\cdots.
 \end{equation}
That is, $\phi=1$ or $0$ in the outer region.

  \subsection{Equation in the inner region}

\begin{figure}
  \centering
  \includegraphics[width=6cm]{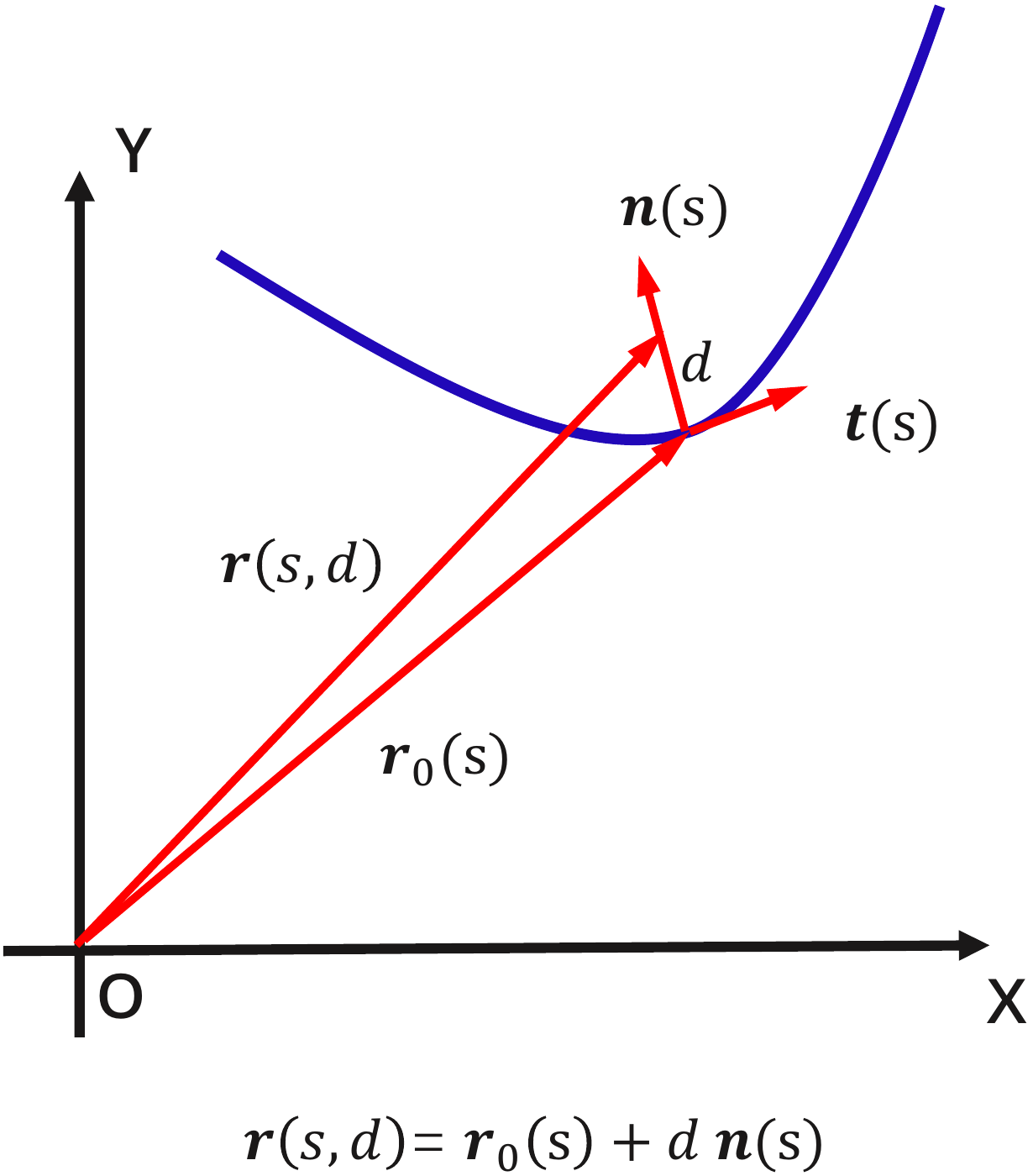}\\
  \caption{Local coordinates $(s,d)$ in the inner region along a dislocation $C$.}\label{fig.localcoordinate}
\end{figure}

  In the inner region, i.e., the small region near the dislocation, we use the local coordinates associated with it. Considering a dislocation $C$, we write it as $\mathbf r_0(s)$  where $s$ is its arclength parameter. Denote that along the dislocation $C$, its unit tangent vector is $\mathbf t(s)$ and unit inward normal vector $\mathbf n(s)$. A point $\mathbf r$ near the dislocation can be written as
  \begin{eqnarray}\label{eqn:r_sd}
  \mathbf r(s,d)=\mathbf r_0(s)+d\mathbf n(s),
  \end{eqnarray}
 where $d$ is the signed distance from the point  $\mathbf r$ to the dislocation. See a schematic illustration in Fig.~\ref{fig.localcoordinate}.


 Since within the dislocation core region, the gradients of fields are of order $1/\varepsilon$, we use variable $\rho=d/\varepsilon$ instead of $d$, and express all the fields using $(s,\rho)$ in the inner region. That is, the two coordinate axes in the local coordinate system is $\mathbf t(s)$ and $\mathbf n(s)$. Note that $\rho>0$ inside the dislocation loop.
From Eq.~\eqref{eqn:r_sd}, we have
\begin{flalign}
\frac{\partial \mathbf r}{\partial s}=&\mathbf t+d\frac{\partial \mathbf n}{\partial s}=(1-\varepsilon\rho\kappa)\mathbf t,\\
\frac{\partial \mathbf r}{\partial \rho}=&\varepsilon\frac{\partial \mathbf r}{\partial d}=\varepsilon\mathbf n,
\end{flalign}
where $\kappa$ is the curvature of the dislocation.

The gradient and divergence operators in the local coordinate system $(s,\rho)$ with coordinate axes $(\mathbf t(s), \mathbf n(s))$ are
  \begin{flalign}\label{eqn.gradient.newsystem}
  \nabla=&\frac{1}{1-\varepsilon\rho\kappa}\textbf{t}\partial_s+\frac{1}{\varepsilon}\textbf{n}\partial_\rho,\\
  \nabla\cdot \mathbf A=&  \frac{1}{1-\varepsilon\rho\kappa}\left(\partial_s A_t+\frac{1}{\varepsilon}\partial_\rho \big((1-\varepsilon\rho\kappa)A_n\big)
  \right),
  \end{flalign}
where $\mathbf A=A_t \mathbf t+A_n \mathbf n$.

In the inner region, we write $\phi(x,y,t)=\Phi(s,\rho,t)$.
The operator $\nabla \cdot (M (\phi)  \nabla)$ in the phase field model in Eq.~\eqref{eqn.phasefield} can be written as
$\nabla \cdot (M (\Phi)  \nabla)= \frac{1}{1-\varepsilon\rho\kappa}\partial_s\left(\frac{1}{1-\varepsilon\rho\kappa} M(\Phi) \partial_s\right)
+\frac{1}{\varepsilon^2}\frac{1}{1-\varepsilon\rho\kappa}\partial_\rho\left( (1-\varepsilon\rho\kappa) M(\Phi) \partial_\rho\right) $
  in this local coordinate system. Note that during the evolution, both $\mathbf t$ and $\mathbf n$ as well as the reference point $\mathbf r_0$ depend on time. In this moving coordinate system, we have
$\phi_t=\Phi_t-\mathbf v\cdot\nabla  \Phi=\Phi_t+\frac{1}{\varepsilon}v_n\partial_\rho \Phi$,
where $\mathbf v$ is the velocity of the dislocation at the point $\mathbf r_0$, and $v_n$ is the normal component of the dislocation velocity.


Therefore, the phase field model in the inner region can be written as
\begin{flalign}\nonumber
\Phi _{t}+\dfrac {1}{\varepsilon}v_{n}\partial _{\rho }\Phi
 =&\dfrac {M_0}{1-\varepsilon\rho\kappa}\partial _{s}\left(\dfrac {1}{1-\varepsilon\rho\kappa} \left( \partial _{s}\mu_c -\dfrac {g' \left( \Phi \right) }{g\left( \Phi \right) }\mu_c \partial _{s}\Phi \right)\right)\\
 &+ \dfrac {1}{\varepsilon^{2}}\dfrac {M_0}{1-\varepsilon\rho\kappa}\partial _{\rho }\left((1-\varepsilon\rho\kappa)\left( \partial _{\rho }\mu_c -\dfrac {g' \left( \Phi \right) }{g\left( \Phi \right) }\mu_c \partial _{\rho }\Phi \right)\right), \label{eqn.inner-eq-1}   \\
       \mu_c=& -\dfrac {1}{1-\varepsilon\rho\kappa}\partial _{s}\left(\dfrac {1}{1-\varepsilon\rho\kappa}\partial _{s}\Phi \right)
       -\dfrac {1}{\varepsilon^{2}}\dfrac {1}{1-\varepsilon\rho\kappa}\partial _{\rho }\left((1-\varepsilon\rho\kappa)\partial _{\rho }\Phi\right)
       \nonumber\\
        &+ \dfrac {1}{\varepsilon^{2}}q'\left( \Phi \right)+\frac{1}{\varepsilon} h\left( \Phi \right)f_{\rm cl}\left(s,\rho,\Phi \right).\label{eqn.inner-eq-mu}
     \end{flalign}

 \subsection{Inner expansion and asymptotic matching}\label{sec:inner}

Based on the local coordinate system given in the previous subsection, we perform expansion of the solution in the inner region (i.e. the solution $\Phi$ of Eqs.~\eqref{eqn.inner-eq-1} and \eqref{eqn.inner-eq-mu}) in terms of the small parameter $\varepsilon$ and asymptotic matching between the solutions in the inner and outer regions. As a result, the sharp interface limit of the dislocation self-climb velocity is determined.

Suppose that the solution in the inner region has the expansion
  \begin{eqnarray}
\Phi \left(s, \rho ,t \right) =\Phi ^{\left( 0\right) }\left( \rho \right) +\varepsilon\Phi ^{\left( 1\right) }\left( s,\rho,t \right) +\varepsilon^{2}\Phi ^{\left( 2\right) }\left( s,\rho,t \right) +\cdots.
  \end{eqnarray}
Especially, we assume that the leading order solution $\Phi ^{\left( 0\right) }\left( \rho\right)$, which describes the dislocation core profile, does not depend on the variable $s$ and $t$, i.e, it is the same at all the points on the dislocation at any time. The chemical potential $\mu_c$ has the same expansion as that in Eq.~\eqref{eqn:mcexpansion}.

 Following Eq.~\eqref{eqn.elasiticity} together with Eqs.~\eqref{eqn:climbforce11} and \eqref{eqn:totalforce}, the climb force $f_{\rm cl}$ in the original coordinate system has the asymptotic property within the dislocation core region:
\begin{flalign}\label{eqn:fcl_inner}
f_{\rm cl}( s,\rho,\Phi)=&\frac{1}{\varepsilon}f_{\rm cl}^{(-1)}( \rho,\Phi^{(0)})+f_{\rm cl}^{(0)}( s)+O(\varepsilon),
\end{flalign}
where
\begin{flalign}
f_{\rm cl}^{(-1)}( \rho,\Phi^{(0)})=&\dfrac {\mu b  }{2\pi \left( 1-\nu \right)}\int^{+\infty}_{-\infty}\frac{\partial _{\rho}\Phi^{(0)}(\rho_1)}{\rho-\rho_1}d\rho_1,\label{eqn:fcl-1}\\
f_{\rm cl}^{(0)}( s)=&f_{\rm cl}^{\rm d}( s)+f_{\rm cl}^{\rm app}( s),\\
f_{\rm cl}^{\rm d}( s)=&\dfrac {\mu b^2}{4\pi \left( 1-\nu \right) }\kappa \ln \varepsilon
+O(1).\label{eqn:fcl-3}
\end{flalign}
Here $\frac{1}{\varepsilon}f_{\rm cl}^{(-1)}( \rho,\Phi^{(0)})$ is due to the singular stress field near the dislocation, which vanishes on the dislocation, i.e., $f_{\rm cl}^{(-1)}(0,\Phi^{(0)})=0$ (guaranteed by the fact that $\Phi^{(0)}(\rho)$ is anti-symmetric at $\rho=0$ about the value $1/2$  due to Eq.~\eqref{eqn.inner-delta mu 0-1}), $f_{\rm cl}^{(0)}( s)$ is the climb force on the dislocation which is the summation of the force generated by dislocations denoted by $f_{\rm cl}^{\rm d}( s)$ and that due to the applied stress denoted by $f_{\rm cl}^{\rm app}( s)$. The climb force  $f_{\rm cl}^{\rm d}( s)$ generated by dislocations has the asymptotic property in Eq.~\eqref{eqn:fcl-3} following Eq.~\eqref{force-appro}. In particular, note that here using the order parameter $\phi$, the dislocation core width is $O(\varepsilon)$, which is used in the asymptotic behavior of the climb force in Eq.~\eqref{eqn:fcl-3}.

 The leading-order of the system of Eqs.~\eqref{eqn.inner-eq-1} and \eqref{eqn.inner-eq-mu} is $O(\varepsilon^{-4})$. The equation on this order is
\begin{flalign}\label{eq-inner-1:4order}
&M_0\partial _{\rho \rho }\mu_c^{\left(0\right) }-M_0\partial _\rho \left(\dfrac {g' \left(\Phi^{\left(0\right) } \right) }{g\left( \Phi^{\left(0\right) } \right) }\mu_c^{\left(0\right) }\partial _{\rho }\Phi ^{\left(0\right)}\right)=0,\\
\label{eqn.inner-delta mu 02}
 & \mu_c^{\left( 0\right) }= -\partial _{\rho \rho }\Phi ^{\left( 0\right) }+q'\left( \Phi ^{\left( 0\right) }\right) +h\left( \Phi^{ \left( 0\right) }\right) f_{\rm{cl}}^{\left( -1\right)} \left(\rho, \Phi ^{\left( 0\right) }\right),
\end{flalign}
where
$f_{\rm{cl}}^{\left( -1\right)} \left(\rho, \Phi ^{\left( 0\right) }\right)$
is given by Eq.~\eqref{eqn:fcl-1} and is independent of $s$ and $t$ because $\Phi^{\left( 0\right) }$ is independent of $s$ and $t$ by assumption.

Eq.~\eqref{eq-inner-1:4order}  can be written as
$\partial _{\rho \rho }\mu_c^{\left( 0\right) }-\partial _{\rho }\left( \mu_c^{\left( 0\right) }\partial _{\rho }\ln g\left(  \Phi^{ \left( 0\right)} \right) \right)=0$.
Integrating both sides of this equation, we have
\begin{eqnarray}\label{eqn.inner-leading order-result-1}
  \partial _{\rho }\mu_c^{\left( 0\right) }-\mu_c^{\left( 0\right) }\partial _{\rho }\ln g\left(  \Phi ^{\left( 0\right) }\right) =C_{1}\left( s\right).
\end{eqnarray}
Since $\mu_c^{\left( 0\right) }=0$ in the outer region, the asymptotic matching gives   $\mu_c^{\left( 0\right) }\rightarrow 0$ and $\partial_\rho \mu_c^{\left( 0\right) }\rightarrow 0$ as   $\rho \rightarrow \pm \infty $. Therefore, we have
$C_{1}\left( s\right) =0$. Dividing both sides of Eq.~\eqref{eqn.inner-leading order-result-1} by $\mu_c^{\left( 0\right) }$ and taking integration, we have
$\ln \big(\mu_c^{ \left( 0\right) }/g\left( \Phi ^{\left( 0\right) }\right)\big)=C_{2}\left( s\right)$, or
  \begin{eqnarray}\label{eqn.inner-1:4w-resulst-2}
\mu_c^{\left( 0\right) }=\tilde {C_{2}}\left( s\right) g\left( \Phi^{ \left( 0\right)} \right).
 \end{eqnarray}

\begin{figure}[htbp]
  \centering
  \includegraphics[width=8cm]{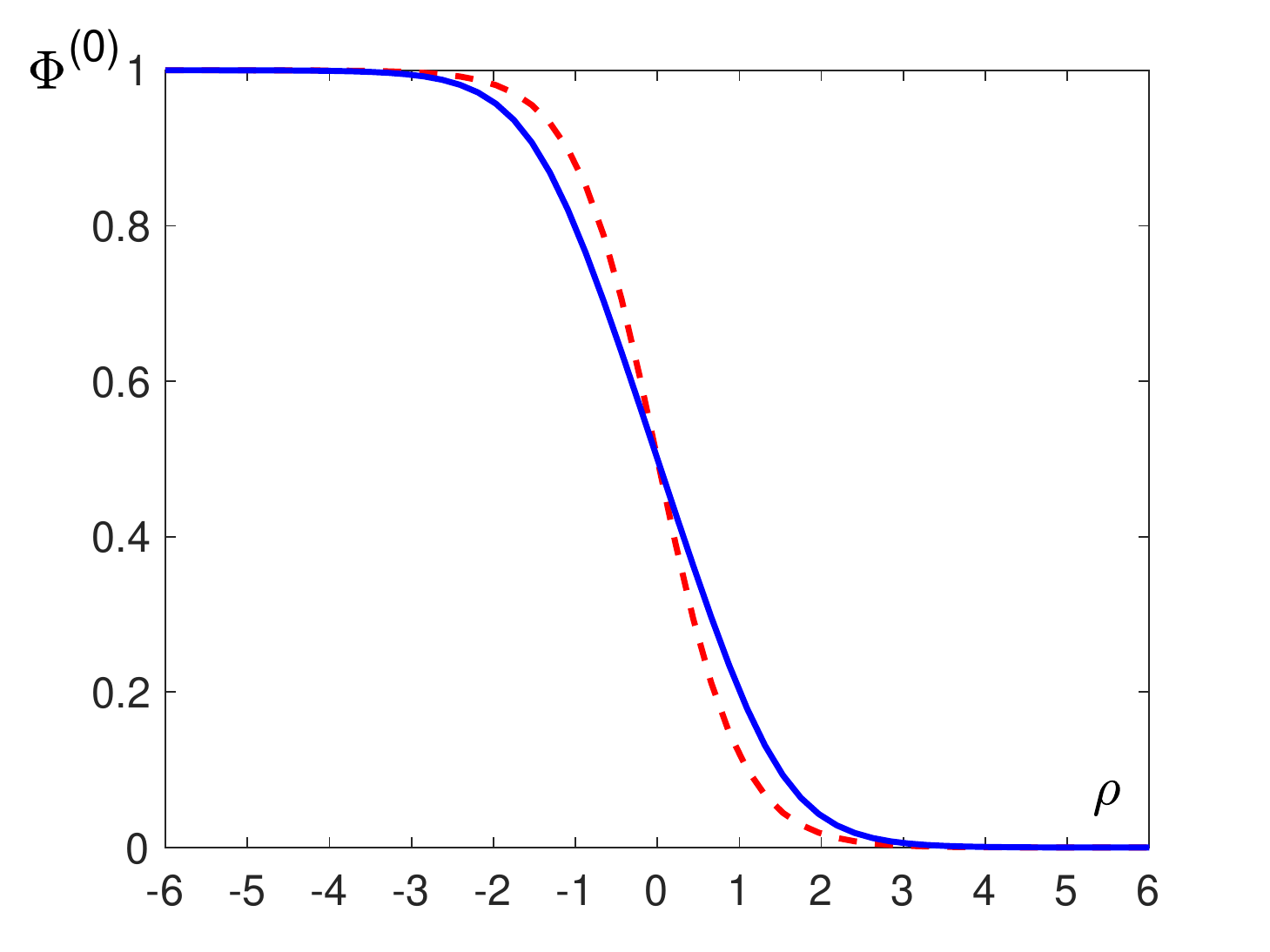}\\
  \caption{Profile of $\Phi^{(0)}(\rho)$ by solving Eq.~\eqref{eqn.inner-delta mu 0-1} with $H_0=52.65\left(2\left( 1-\nu \right) /\mu b^2\right)$ in $h(\phi)$ (solid blue line), and comparison with that in the classical Cahn-Hilliard equation (dashed red line).}\label{fig.core}
\end{figure}

Since $\Phi^{\left( 0\right) }$ is independent of $s$,
 $\mu_c^{\left( 0\right) }$ given by Eq.~\eqref{eqn.inner-delta mu 02} is also independent of $s$, Eq.~\eqref{eqn.inner-1:4w-resulst-2} holds only when
$\tilde  {C}_{2}\left( s\right)=0$.  Thus we have
\begin{eqnarray} \label{eqn.inner-delta mu 0-1}
 \mu_c^{(0) }=-\partial _{\rho \rho }\Phi ^{\left( 0\right) }+q'\left( \Phi ^{\left( 0\right) }\right) +h\left( \Phi^{ \left( 0\right) }\right) f_{\rm{cl}}^{\left( -1\right)} \left(\rho, \Phi ^{\left( 0\right) }\right)=0,
\end{eqnarray}
where $f_{\rm{cl}}^{\left( -1\right)} \left(\rho, \Phi ^{\left( 0\right) }\right)$ is given by Eq.\eqref{eqn:fcl-1}. The solution $\Phi^{\left( 0\right) }$ of this equation with the far field conditions
$\Phi^{\left( 0\right) }(+\infty)=0$ and $\Phi^{\left( 0\right) }(-\infty)=1$ gives the time-independent dislocation core profile, and can be obtained numerically by solving the gradient flow equation $\Phi_{\tau}^{\left( 0\right) }=  \partial _{\rho \rho }\Phi ^{\left( 0\right) }-q'\left( \Phi ^{\left( 0\right) }\right) -h\left( \Phi^{ \left( 0\right) }\right) f_{\rm{cl}}^{\left( -1\right)} \left(\rho, \Phi ^{\left( 0\right) }\right)$ to equilibrium, where $\tau$ is some artificial time. The obtained profile of $\Phi ^{\left( 0\right) }(\rho)$ is shown in Fig.~\ref{fig.core}, with the parameter $H_0=52.65\frac {2\left( 1-\nu \right) }{\mu b^2}$ in $h(\phi)$ used in the simulations in Sec.~\ref{sec:simulation}. It can be seen that the half-width of the interface (dislocation core) is about $1.5$, which only slightly change the one in the classical Cahn-Hilliard equation (i.e., without the climb force term in Eq.~\eqref{eqn.inner-delta mu 0-1}) whose value is about $1$. This means that in the phase field model before rescaling by $\varepsilon$, the half-width of the interface (dislocation core) is about $1.75\varepsilon$ whereas that in the classical Cahn-Hilliard equation  is about $\varepsilon$.

Next, we consider the  $O(\varepsilon^{-3})$ problem of the system of Eqs.~\eqref{eqn.inner-eq-1} and \eqref{eqn.inner-eq-mu}. Using
 $\mu_c^{(0) }=0$, it can be calculated that
 the  $O(\varepsilon^{-3})$ equation is
\begin{flalign}\label{eqn.inner-next-to-leading}
&\partial_{ \rho \rho}\mu_c^{\left(1\right)}-\partial_ \rho\left (\frac{g' \left(\Phi^{ \left( 0\right) } \right)}{g\left( \Phi^{ \left( 0\right) } \right)}\mu_c^{\left(1\right)}\partial _\rho \Phi^{ \left(0\right)}\right)=0,
\end{flalign}
where
\begin{flalign}\label{eq.inner-delta mu 1}
&\mu_c^{\left( 1\right) }=-\partial _{\rho \rho }\Phi^{  \left( 1\right)} +\kappa\partial _{\rho }\Phi ^{ \left( 0\right) }+q^{''}\left( \Phi ^{ \left( 0\right)} \right) \Phi ^{ \left( 1\right)}
+h'\left( \Phi^{ \left( 0\right) }\right) f_{\rm{cl}} ^{(-1)}\left( \rho,\Phi^{  \left( 0\right) }\right)\Phi^{  \left( 1\right)}
+h\left( \Phi^{  \left( 0\right)} \right) f_{\rm{cl}}^{(0)} \left(s\right).
\end{flalign}
Here we have used the asymptotic expansion in Eq.~\eqref{eqn:fcl_inner}.
Integrating both sides of Eq.~\eqref{eqn.inner-next-to-leading} gives
$ \partial _{\rho }\mu_c^{\left( 1\right) }-\mu_c^{\left( 1\right) }\partial _{\rho }\ln g\left(   \Phi ^{\left( 0\right) }\right) =C_{3}\left( s\right)$.
Matching the inner solutions with the corresponding outer solutions, we have
$\mu_c^{\left( 1\right) }, \partial _{\rho }\mu_c^{\left( 1\right) }\rightarrow 0
$ as $\rho \rightarrow \pm \infty $. Thus we have $C_{3}\left( s\right) =0$, and 
 $\partial _{\rho }\mu_c^{\left( 1\right) }-\mu_c^{\left( 1\right) }\partial _{\rho }\ln g\left(   \Phi ^{\left( 0\right) }\right)=0$, or
                     $\partial _{\rho}\ln \left(\mu_c^{\left( 1\right) }/g\left(  \Phi ^{\left( 0\right) }\right) \right)=0$.
Thus, we have
\begin{equation}\label{eqn:mc1ds}
\dfrac {\mu_c^{\left( 1\right) }}{g\left(  \Phi ^{\left( 0\right) }\right) }=D_{1}\left( s\right).
\end{equation}
Using the expression of $\mu_c^{\left( 1\right) }$ in Eq.~\eqref{eq.inner-delta mu 1}, we have
 \begin{flalign}\nonumber
 &-\partial _{\rho \rho }\Phi^{  \left( 1\right)} +\kappa\partial _{\rho }\Phi ^{ \left( 0\right) }+q^{''}\left( \Phi ^{ \left( 0\right)} \right) \Phi ^{ \left( 1\right)}
+h'\left( \Phi^{ \left( 0\right) }\right) f_{\rm{cl}} ^{(-1)}\left( \rho,\Phi^{  \left( 0\right) }\right)\Phi^{  \left( 1\right)}
+h\left( \Phi^{  \left( 0\right)} \right) f_{\rm{cl}}^{(0)} \left(s\right)\\
=&D_{1}\left( s\right)g\left(  \Phi ^{\left( 0\right) }\right).\label{eqn:mu1}
\end{flalign}
Note that Eq.~\eqref{eqn:mu1} satisfies the matching conditions as $\rho\rightarrow \pm\infty$ even though $D_{1}\left( s\right)$ is not identical to $0$.

We multiply Eq.~\eqref{eqn:mu1} by $\partial_\rho \Phi ^{\left( 0\right) }$ and  integrate  with respect to $\rho$ over $(-\infty, +\infty)$.  The terms that contain  $\Phi^{  \left( 1\right)}$ in the resulting equation are
\begin{flalign}\nonumber
&\int_{-\infty}^{+\infty}\left( -\partial _{\rho \rho }\Phi^{  \left( 1\right)}
+q^{''}\left( \Phi ^{ \left( 0\right)} \right) \Phi ^{ \left( 1\right)}
+h'\left( \Phi^{ \left( 0\right) }\right) f_{\rm{cl}} ^{(-1)}\left( \rho,\Phi^{  \left( 0\right) }\right)\Phi^{  \left( 1\right)}
\right) \partial_\rho \Phi ^{\left( 0\right) }d\rho\\
  =&\int_{-\infty}^{+\infty}\left( \partial_{\rho\rho}\Phi ^{\left( 0\right) }- q^{'}\left( \Phi ^{ \left( 0\right)}\right) - h\left( \Phi^{ \left( 0\right) }\right)  f_{\rm{cl}} ^{-1}\left(\rho,\Phi ^{ \left( 0\right)}\right)\right)\partial_\rho \Phi ^{ \left( 1\right)}d\rho.
  \end{flalign}
Following  Eq.~\eqref{eqn.inner-delta mu 0-1}, this integral vanishes. The remaining terms in the resulting equation are
\begin{equation}
\int_{-\infty}^{+\infty}\left(   \kappa\partial _{\rho }\Phi ^{ \left( 0\right) }+h\left( \Phi^{  \left( 0\right)} \right) f_{\rm{cl}}^{(0)} \left( s\right)\right)\partial_\rho \Phi ^{ \left( 0\right)}d\rho
=\int_{-\infty}^{+\infty}D_{1}\left( s\right)g\left(  \Phi ^{\left( 0\right) }\right)\partial_\rho \Phi ^{ \left( 0\right)}d\rho.
 \end{equation}
Thus, we have
\begin{eqnarray}\label{eqn:D1}
 D_{1}\left( s\right) =-\alpha_0\kappa+H_0 f_{\rm{cl}}^{(0)}\left(  s \right) ,
\end{eqnarray}
where
\begin{equation}\label{eqn:alpha_0}
 \alpha_0=-\frac{\int_{-\infty}^{+\infty}(\partial_\rho \Phi ^{\left( 0\right) } )^2d \rho}{\int_{-\infty}^{+\infty}g\left(  \Phi ^{\left( 0\right) }\right)\partial_\rho \Phi ^{\left( 0\right) }d\rho}.
 \end{equation}
Recall that $h\left( \Phi^{  \left( 0\right)} \right)=H_0q\left( \Phi^{  \left( 0\right)} \right)=H_0g\left( \Phi^{  \left( 0\right)} \right)$.
Note that $\alpha_0>0$ due to $\partial_\rho \Phi ^{\left( 0\right) }<0$.
Therefore, we have
\begin{equation}
\mu_c^{\left( 1\right) }=g\left(  \Phi ^{\left( 0\right) }\right)\left(- \alpha_0\kappa+ H_0 f_{\rm{cl}}^{(0)}\left( s \right)\right).
\end{equation}

Now we consider the  $O(\varepsilon^{-2})$ problem. Denoting $\bar {\mu }=\dfrac{ \mu_c }{g\left(  \Phi \right)} $,  the inner equation \eqref{eqn.inner-eq-1}  becomes
\begin{equation}\label{eqn.inner-eq-12}
\Phi _{t}+\dfrac {1}{\varepsilon}v_{n}\partial _{\rho }\Phi = \dfrac {M_0}{1-\varepsilon \rho\kappa}\partial _{s}\left(\dfrac {1}{1-\varepsilon \rho\kappa}g\left(\Phi\right) \partial _{s}\bar{\mu}\right)
+\dfrac {1}{\varepsilon^{2}}\dfrac {M_0}{1-\varepsilon \rho\kappa}\partial _{\rho }\left((1-\varepsilon \rho\kappa) g \left(\Phi\right) \partial _{\rho }\bar {\mu}\right).
\end{equation}
From Eqs.~\eqref{eqn.inner-delta mu 0-1} and \eqref{eqn:mc1ds},
 $\mu_c^{\left( 0\right) }=0$, $ \partial _{\rho }\left( \frac {\mu_c^{\left( 1\right) }}{g\left(  \Phi ^{\left( 0\right) }\right) }\right)=0$, thus we have
 $\bar {\mu}^{\left( 0\right) }=\frac {\mu_c^{\left( 0\right) }}{g\left(  \Phi ^{\left( 0\right) }\right) }=0$
  and $ \partial_\rho \bar {\mu}^{\left( 1\right) }=\partial _{\rho }\left( \frac {\mu_c^{\left( 1\right) }}{g\left(  \Phi ^{\left( 0\right) }\right) }\right)=0$,
 The $O(\varepsilon^{-2})$ equation of Eq.~\eqref{eqn.inner-eq-12} is
\begin{eqnarray}
 \partial _{\rho }\left(g\left( \Phi ^{\left( 0\right) }\right) \partial _{\rho } \bar  {\mu }^{\left( 2\right) }\right)=0.
\end{eqnarray}
  This gives $g\left( \Phi ^{\left( 0\right) }\right) \partial _{\rho }\bar {\mu}^{\left( 2\right) }=C_{4}\left( s\right)$. Since  the left-hand side of this equation goes to zero as $\rho\rightarrow \pm\infty$, we have $C_{4}\left( s\right) =0$ and
  \begin{equation}
   \bar {\mu}^{\left( 2\right) }= \left(\dfrac{ \mu_c }{g\left(  \Phi \right)}\right)^{\left( 2\right) }=D_{2}\left( s\right).
   \end{equation}

The $O(\varepsilon^{-1})$ equation, using Eq.~\eqref{eqn.inner-eq-12} and
 $\varepsilon^{-2}$, using  $ \bar {\mu}^{\left( 0\right) }=0$, $\partial _{\rho }\bar {\mu}^{\left( 1\right) } =0$ and $\partial _{\rho }\bar {\mu}^{\left( 2\right) } =0$, is
\begin{eqnarray}
  v_{n}\partial _{\rho }\Phi ^{\left( 0\right)} =M_0\partial _{\rho }\left(g\left( \Phi ^{\left( 0\right)} \right) \partial _{\rho } \bar {\mu}^{\left( 3\right) }\right)
  +M_0\partial _{s}\left(g\left( \Phi ^{\left( 0\right) }\right) \partial _{s}\bar {\mu}^{\left( 1\right) }\right).
 \end{eqnarray}
 Integrating with respect to $\rho$ over $(-\infty,+\infty)$ on both sides of this equation, and noticing that $\Phi ^{\left( 0\right)} =\Phi ^{\left( 0\right)} \left( \rho\right)$, we have
 \begin{equation}
-v_{n}=M_0\partial _{ss} \bar {\mu }^{\left( 1\right) }\int ^{+\infty }_{-\infty }g\left( \Phi^{ \left( 0\right)} (\rho)\right) d\rho.
 \end{equation}
 Using the definition of $\bar{\mu}$ and Eqs.~\eqref{eqn:mc1ds} and \eqref{eqn:D1}, we have
 $ \bar {\mu}^{\left( 1\right) }= \frac {\mu_c^{\left( 1\right) }}{g\left(  \Phi ^{\left( 0\right) }\right) }=D_1(s)=-\alpha_0\kappa+H_0 f_{\rm{cl}}^{(0)}\left(  s \right)$, and
  \begin{equation}\label{eqn:sharplimit}
    v _{n}
    =-M_0\alpha H_0\partial _{ss}\left(-\frac{\alpha_{0}}{H_0}\kappa+f_{\rm{cl}}^{(0)}\left(  s \right) \right),
 \end{equation}
where
\begin{equation}\label{eqn:alpha}
\alpha=\int ^{+\infty }_{-\infty }g\left( \Phi^{ \left( 0\right)} (\rho)\right)  d\rho.
\end{equation}
Note that in Eq.~\eqref{eqn:sharplimit}, both $v _{n}$ and $f_{\rm{cl}}^{(0)}$ are in the $\mathbf l_{\rm cl}$ direction defined in Eq.~\eqref{eqn:climbdirection}, which is in the outer normal direction of the dislocation loop (vacancy loop).

Eq.~\eqref{eqn:sharplimit} is the sharp interface limit of the dislocation self-climb velocity. Compared with the dislocation self-climb velocity formula in Eq.~\eqref{eqn.self-climb-v}, in addition to the dislocation self-climb force $f_{\rm{cl}}^{(0)}$, there is an extra term of dislocation curvature generated by the framework of the phase field model. This extra curvature term serves as a corrector, instead of an error, to the numerical formulation of the climb force $f_{\rm{cl}}^{\rm d}$ in Eq.~\eqref{eqn:fcl_pf} in terms of the order parameter $\phi$. In fact,  the numerical dislocation core in the phase field model is given by the small parameter $\varepsilon$, which in general is of $O(10b)$, is always larger than its actual size $r_d$ of $O(b)$. The difference between the asymptotic behavior in Eq.~\eqref{eqn:fcl-3} of the climb force calculated by the phase field model and that in the exact formula in Eq.~\eqref{force-appro} is $\frac {\mu b^2}{4\pi \left( 1-\nu \right) }\kappa \ln \frac{\varepsilon}{r_d}$. Therefore, if we choose $\alpha_0/H_0=\frac {\mu b^2}{4\pi \left( 1-\nu \right) } \ln \frac{\varepsilon}{r_d}$, the climb force calculated in the phase field model will become more accurate. This gives
\begin{equation}\label{eqn:H0}
H_0=\frac{\alpha_0}{\frac {1}{2\pi  }\ln \frac{\varepsilon}{r_d}}\cdot \frac {2\left( 1-\nu \right) }{\mu b^2}.
\end{equation}
Accordingly, comparing the coefficients in Eqs.~\eqref{eqn:sharplimit} and \eqref{eqn.self-climb-v}, we set the constant $M_0$ in the mobility of the phase field model to be
\begin{equation}\label{eqn:M0}
M_0=\frac {1}{2\pi \alpha \alpha_0 }\ln \frac{\varepsilon}{r_d}\cdot\frac{\mu b^2 c_0D_c\Omega}{2(1-\nu)k_B T}.
\end{equation}
Note that here constants $\alpha_0$ and $\alpha$ are given in Eqs.~\eqref{eqn:alpha_0} and \eqref{eqn:alpha}.
With these parameters, the phase field model will give an accurate dislocation self velocity as that in Eq.~\eqref{eqn.self-climb-v}.

\subsection{Summary of the phase field model}

We have developed a phase field model for the self-climb motion of dislocations  in Eqs.~\eqref{eqn.phasefield} and \eqref{eqn:mu_c}, where the climb force $f_{\rm cl}$ calculated by Eqs.~\eqref{eqn:totalforce} and \eqref{eqn:fcl_pf}  is incorporated in the classical Cahn-Hilliard equation.
The functions $q(\phi)=2\phi^2(1-\phi)^2$, $g(\phi)=\phi^2(1-\phi)^2$, $h(\phi)=H_0g(\phi)$, $M(\phi)=M_0g(\phi)$, and parameters $H_0$ and $M_0$ are given in Eqs.~\eqref{eqn:H0} and \eqref{eqn:M0} together with $\alpha_0$ and $\alpha$ in Eqs.~\eqref{eqn:alpha_0} and \eqref{eqn:alpha}.

This phase field model is able to generate the dislocation self-climb velocity in Eq.~\eqref{eqn.self-climb-v}. Incorporation of the climb force $f_{\rm cl}$ in the framework of the classical Cahn-Hilliard equation  does not lead to significant changes in the interface profile  (dislocation core profile here) of  the classical Cahn-Hilliard phase field model, and the framework of the phase field model gives an extra term that serves to correct the dislocation climb force calculated by the phase field formulation.

 \section{Simulations}\label{sec:simulation}

 In this section, we will use the proposed phase field model in Eqs.~\eqref{eqn.phasefield} and \eqref{eqn:mu_c}, to simulate the self-climb motion of prismatic dislocation loops. Recall that an important property of self-climb motion is conservation of the enclosed area of a prismatic loop \cite{Kroupa1961}, which is guaranteed by the self-climb formulation in Eq.~\eqref{eqn.self-climb-v}~\cite{Niu2017,Niu2019}. In the simulations, we choose the simulation domain $  \Omega = [-\pi, \pi]^2$ and mesh size $dx=dy=2\pi/M$ with $M=64$. Periodic boundary conditions are used for the simulation domain. The small parameter in the phase field model $\varepsilon=dx$. The simulation domain corresponds to a physical domain of size $(300b)^2$, i.e., $b=2\pi/300$. Under this setting, the parameter $H_0$ in the phase field model calculated by Eq.~\eqref{eqn:H0} is $H_0=52.65\left(2( 1-\nu ) /\mu b^2\right)$. The prismatic loops are in the counterclockwise direction meaning vacancy loops, unless otherwise specified.

In the numerical simulations, we use the pseudospectral method: All the spatial partial derivatives are calculated in the Fourier space using FFT. For the time discretization, we use  the forward Euler method. The climb force generated by dislocations $f_{\rm{cl}}^{\rm d}$ is calculated by FFT using Eq.~\eqref{eqn:fftfcl}.
We regularize the function $g(\phi)$ in the denominator in Eq.~\eqref{eqn.phasefield} as $\sqrt{g(\phi)^2+e_0^2}$ with small parameter $e_0=0.005$. In the initial configuration of a simulation, $\phi$ in the dislocation core region is set to be a $\tanh$ function with width $3\varepsilon$. The location of the dislocation loop is identified by the contour line of $\phi=0.5$.

\subsection{Evolution of an elliptic prismatic loop}

\begin{figure}[h]
  \centering
  \includegraphics[width=6cm]{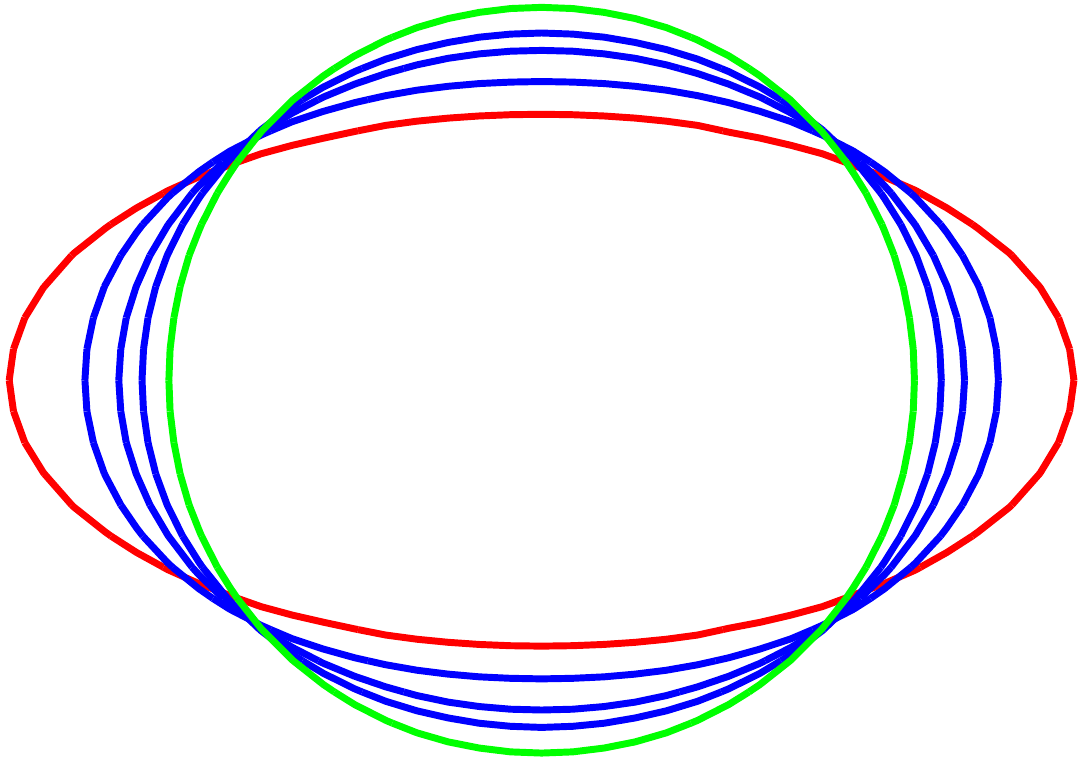}\\
  \caption{Evolution of an elliptic prismatic loop by self-climb  using the phase field model. Red ellipse is the initial state, and green circle is the final state.}\label{fig.elliptic}
\end{figure}

In the first numerical example, we simulate evolution of an elliptic prismatic loop using the phase field model; see Fig.~\ref{fig.elliptic}. The two axes of the initial elliptic profile are $l_1=80b$ and $l_2=40b$.
It can be seen that the prismatic loop evolves into a stable circle.  The radius of the  stable circle is $54.9b$. The theoretical value of radius of the circle loop under conservation of the enclosed area is  $R=\sqrt{l_1l_2}=56.6b$.  Agreement can be seen from the phase field simulation and the experimental~\cite{Hirth-Lothe,Kroupa1961} and theoretical~\cite{Niu2017} predictions that the area enclosed by a prismatic loop is conserved during the self-climb motion and the loop converges to the equilibrium shape of a circle under its self-stress.

In the DDD simulation with the same setting performed in Ref.~\cite{Niu2019}, the radius of the converged circular loop is $R=56.1b$, and the time for the evolution from the initial elliptic loop to the final circular loop is about $t=1.13\times 10^7 \left(\frac{1}{2(1-\nu)}\frac{\mu \Omega}{k_BT}\frac{c_0D_c}{b^2}\right)^{-1}$. In the phase field simulation, this evolution time is  $t=1.21\times 10^7 \left(\frac{1}{2(1-\nu)}\frac{\mu \Omega}{k_BT}\frac{c_0D_c}{b^2}\right)^{-1}$.
 Excellent agreement can be seen between the simulation results using the developed phase field model and those of the DDD simulation.

\subsection{Translation of a circular prismatic loop under constant stress gradient}

In this subsection, we perform simulation using the phase field model on translation of a circular prismatic dislocation loop under constant stress gradient.  The initial prismatic loop is a circle with  radius  $R=100b$.   The applied stress field is $\sigma_{33}^{\rm app}=-px$ with $p=10^{-5}\mu/b$.

\begin{figure}[h]
  \centering
  \includegraphics[width=5cm]{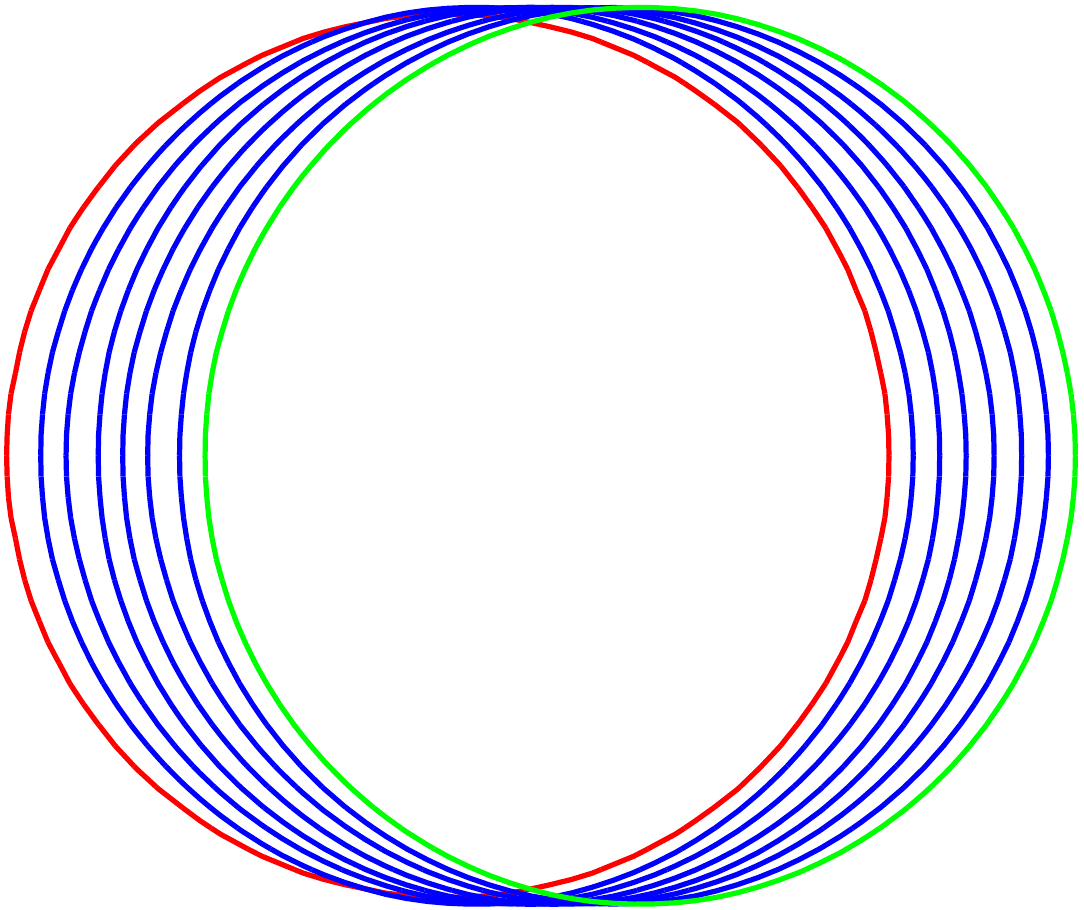}\\
  \caption{Translation of a circular prismatic loop under constant stress gradient using the phase field model. The leftmost, red circle is the initial configuration of the loop.}\label{fig.CircleTranslation}
\end{figure}

As shown in Fig.~\ref{fig.CircleTranslation},
 under the constant stress gradient field, the circular prismatic loop moves with a constant translation velocity while remaining  its circular shape. This result agrees with the theoretical prediction in Refs.~\cite{Niu2017,Niu2019}  based on the climb velocity formula in Eq.~\eqref{eqn.self-climb-v} and the DDD simulation of the same prismatic loop in Ref.~\cite{Niu2019}. Quantitatively, the translation velocity of the loop estimated from the phase field simulation is  $v=1.95\times 10^{-5}c_0D_c/b$. This value is in excellent agreement with the DDD value $v=1.94\times 10^{-5}c_0D_c/b$ in Ref.~\cite{Niu2019} and the theoretical value  $v=1.85\times 10^{-5}c_0D_c/b$ given by  Eq.~(4.16) in Ref.~\cite{Niu2019} for the translation velocity formula of a circular loop.
 This behavior of translation of prismatic loops under stress gradient also agrees with the experimental observations and atomistic simulations~\cite{Hirth-Lothe,Kroupa1961,Turnbull1970,Narayan1972,Swinburne2016,Okita2016}  and the results
 using velocity formulas based on the assumption of rigid translation of a circular prismatic loop~\cite{Johnson1960,Turnbull1970,Narayan1972,Swinburne2016,Okita2016}.

Note that in the front-tracking DDD simulation in Ref.~\cite{Niu2019},   the numerical nodal points are moving in the opposite direction with respect to the loop motion direction and tend to be accumulated in the back side of the loop, which requires remeshing regularly during the evolution. Such remeshing is not needed when using the proposed phase field model.

   \subsection{Coalescence of prismatic loops}

\begin{figure}[htbp]
\centering
\subfigure[]{\includegraphics[width=2.in]{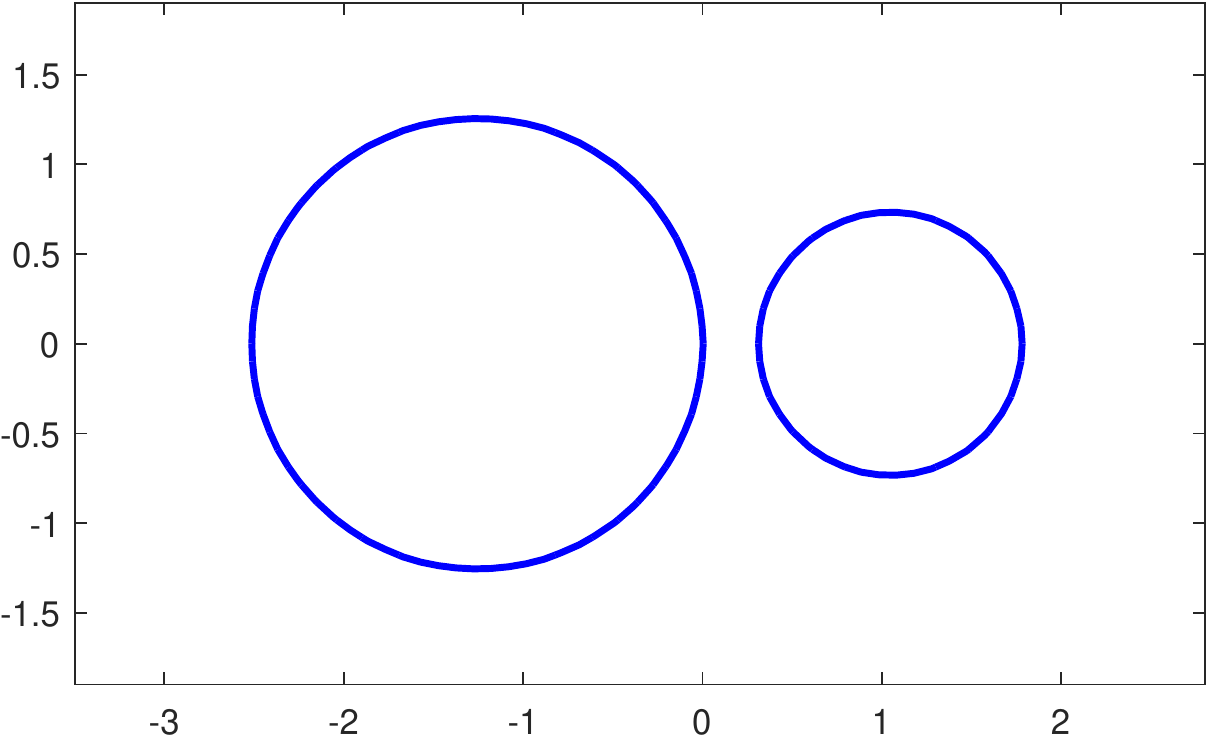}}
\subfigure[]{\includegraphics[width=2.in]{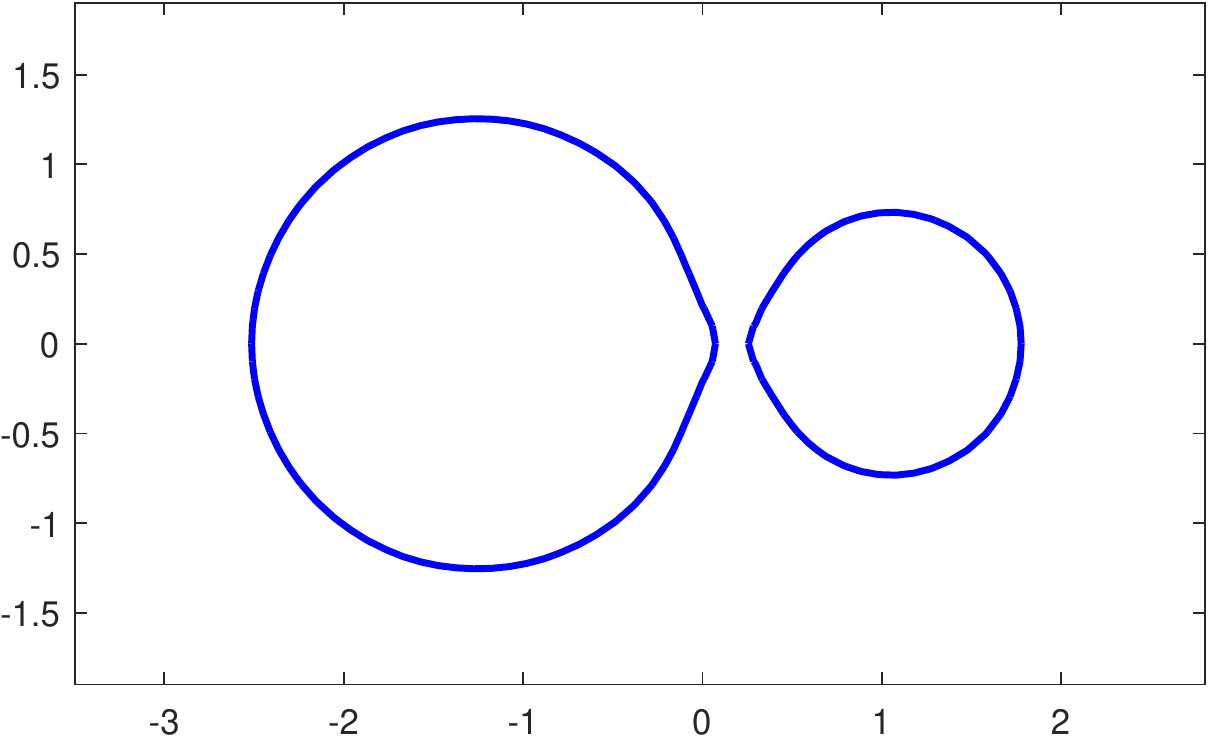}}
\subfigure[]{\includegraphics[width=2.in]{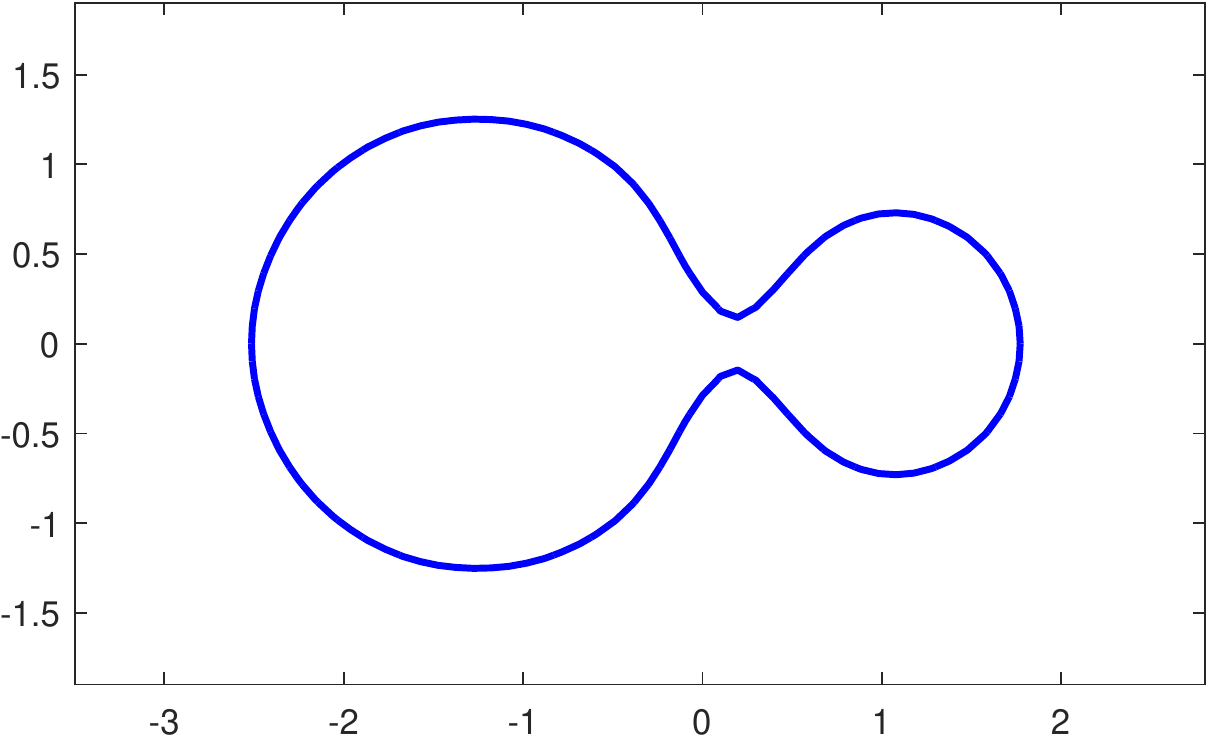}}
\subfigure[]{\includegraphics[width=2.in]{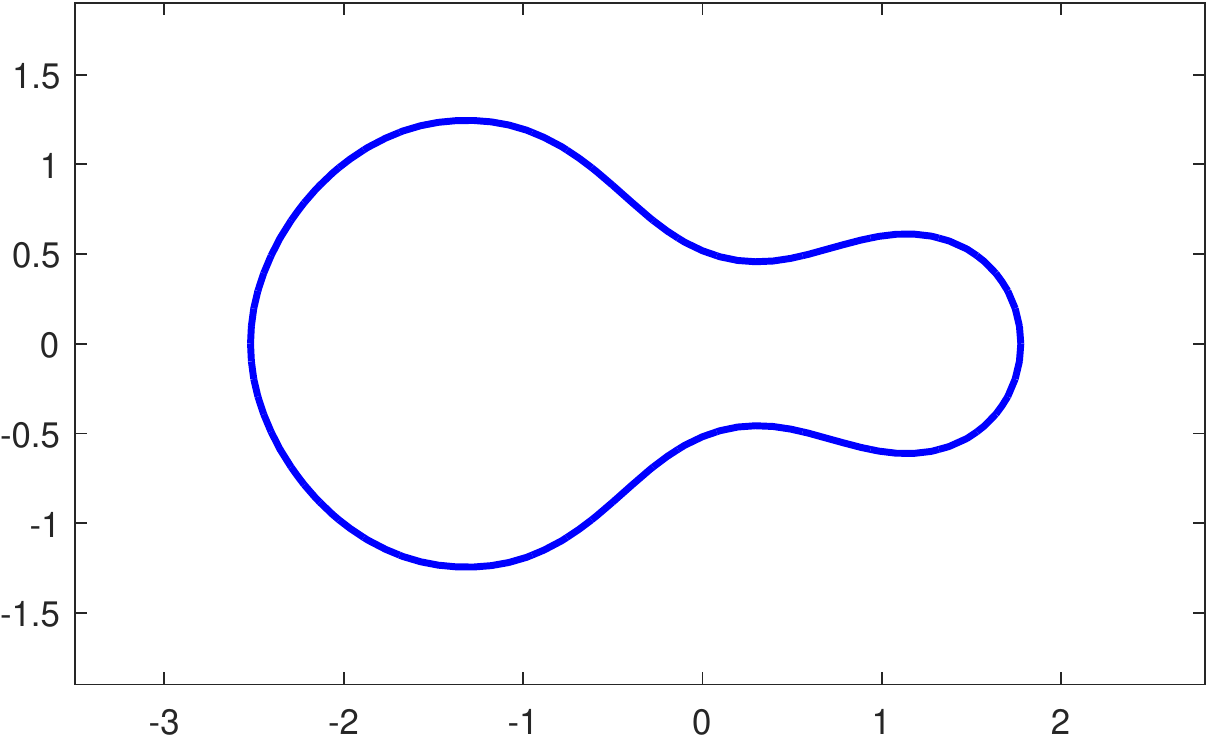}}
\subfigure[]{\includegraphics[width=2.in]{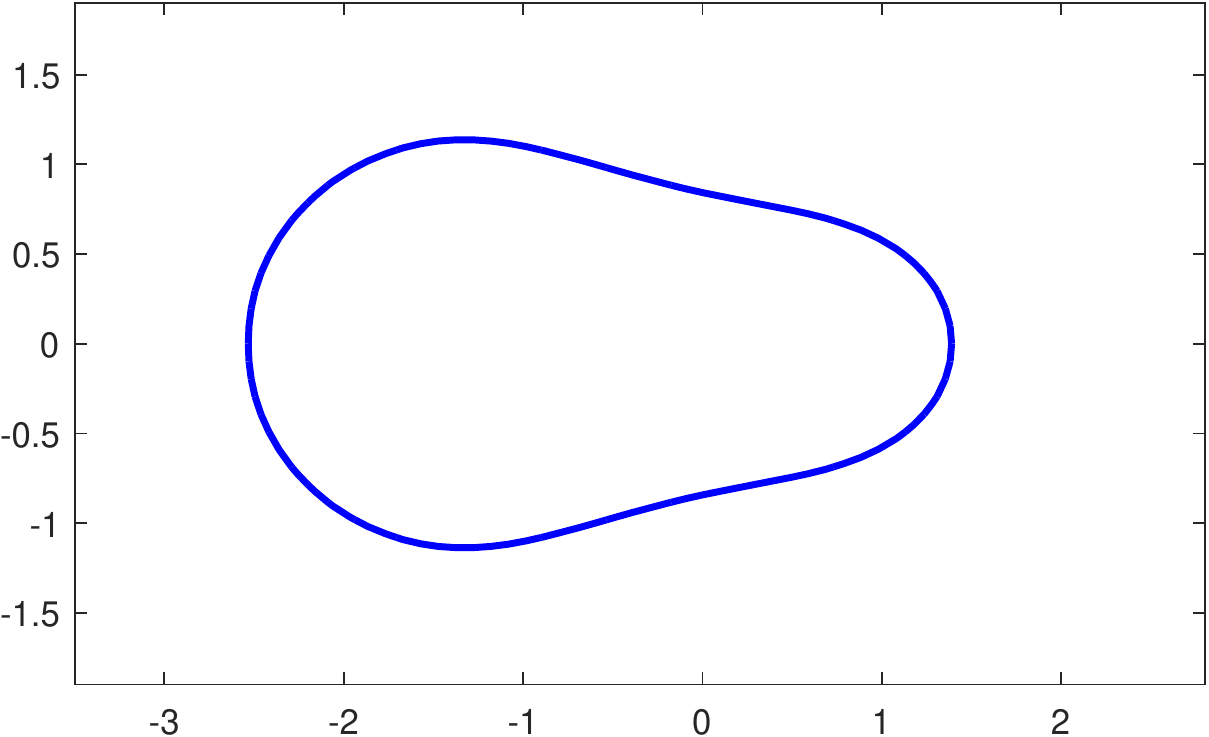}}
\subfigure[]{\includegraphics[width=2.in]{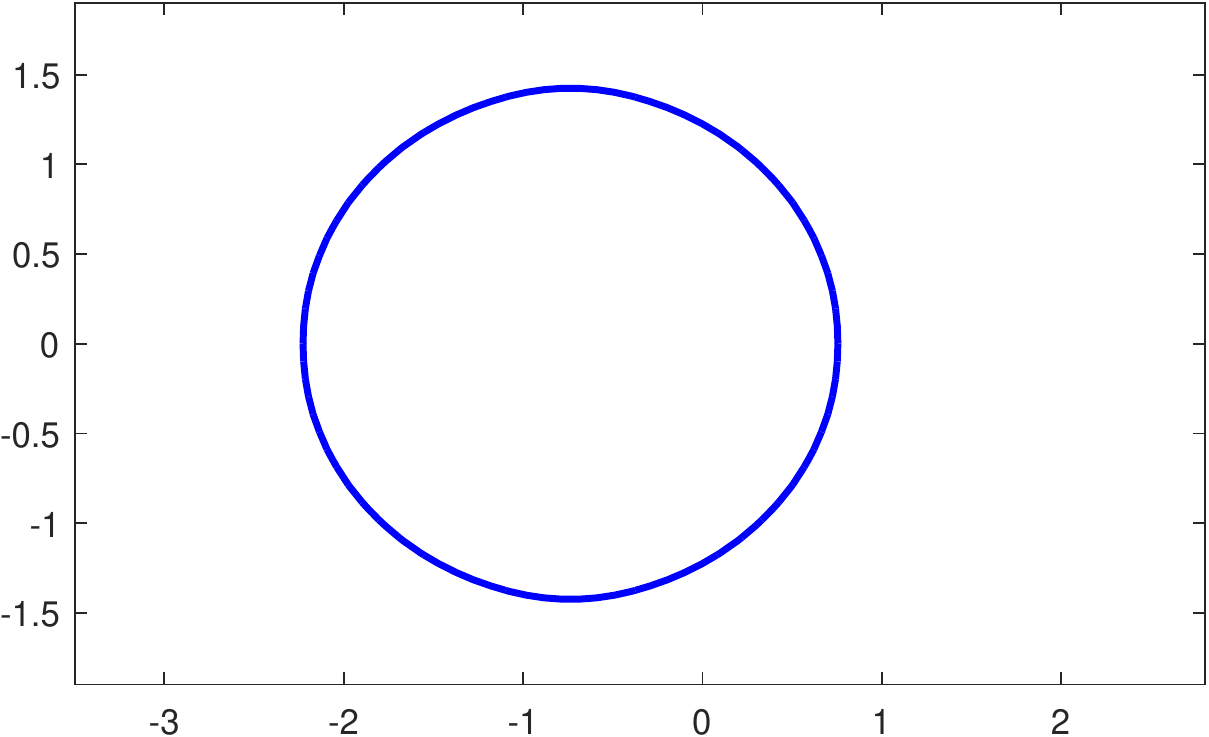}}
	\caption{Coalescence of two prismatic loops by self-climb under their elastic interaction obtained by the phase field model.}\label{fig.twoloopsattractive1}
\end{figure}

  In this subsection, we use our  phase field model to simulate  the coalescence of prismatic loops by self-climb that has been observed in experiments \cite{Silcox1960,Washburn1960,Johnson1960,Turnbull1970,Narayan1972,Swinburne2016} and has been simulated by front-tracking DDD method in Ref.~\cite{Niu2019}.
 We consider two circular prismatic loops as shown in Fig.~\ref{fig.twoloopsattractive1}(a). The two loop have radii of $R_1=60b$ and $R_2=35b$, respectively, and are separated (from center to center) by $d=110b$.

 The detailed coalescence process obtained by our simulation is shown in Fig.~\ref{fig.twoloopsattractive1}. The two loops are attracted to each other by self-climb under the elastic interaction between them,  and the two loops deviate from the circular shape; see Fig.~\ref{fig.twoloopsattractive1}(b).   When the two loops meet, they quickly combine into a single loop; see Fig.~\ref{fig.twoloopsattractive1}(c). The combined single loop eventually evolves into a stable, circular shape; see Fig.~\ref{fig.twoloopsattractive1}(d)-(f). The theoretical value of the final single circular loop under conservation of the total enclosed area has radius $R=\sqrt{R_1^2+R_2^2}=69.5b$. The radius of the final single circular loop obtained by the phase field simulation is about $R=69.6b$, which agrees perfectly with the theoretical value. This simulation result also agrees with the front-tracking DDD simulation result \cite{Niu2019} and experimental observations \cite{Silcox1960,Washburn1960,Johnson1960,Turnbull1970,Narayan1972,Swinburne2016}.

  In the front-tracking DDD simulations performed in Ref.~\cite{Niu2019},  during the  coalescence of two prismatic loops, extremely small time step is needed to avoid numerical instability when two segments from different loops are very close to each other, and at some moment, a decision has to be made to connect these two segments manually to complete the coalescence process. In the phase field simulation, this topological change is handled automatically without extra numerical treatment.

We further perform a simulation using our phase field model for the coalescence of seven circular prismatic loops  by self-climb under their elastic interaction. These circular loops have
     radii $15b$, $20b$, $25b$, $30b$, $35b$, $40b$ and $40b$, respectively, and are placed randomly in the simulation domain. The initial configuration and the  coalescence process are shown in
    Fig.~\ref{fig.7loops}. These loops are coalesced into a
    stable circle with radius $ 82.2b$, which agrees well with the theoretical value $ 81.9 b$ under conservation of the total enclosed area.

    \begin{figure}[htbp]
\centering
\subfigure[]{\includegraphics[width=1.8in]{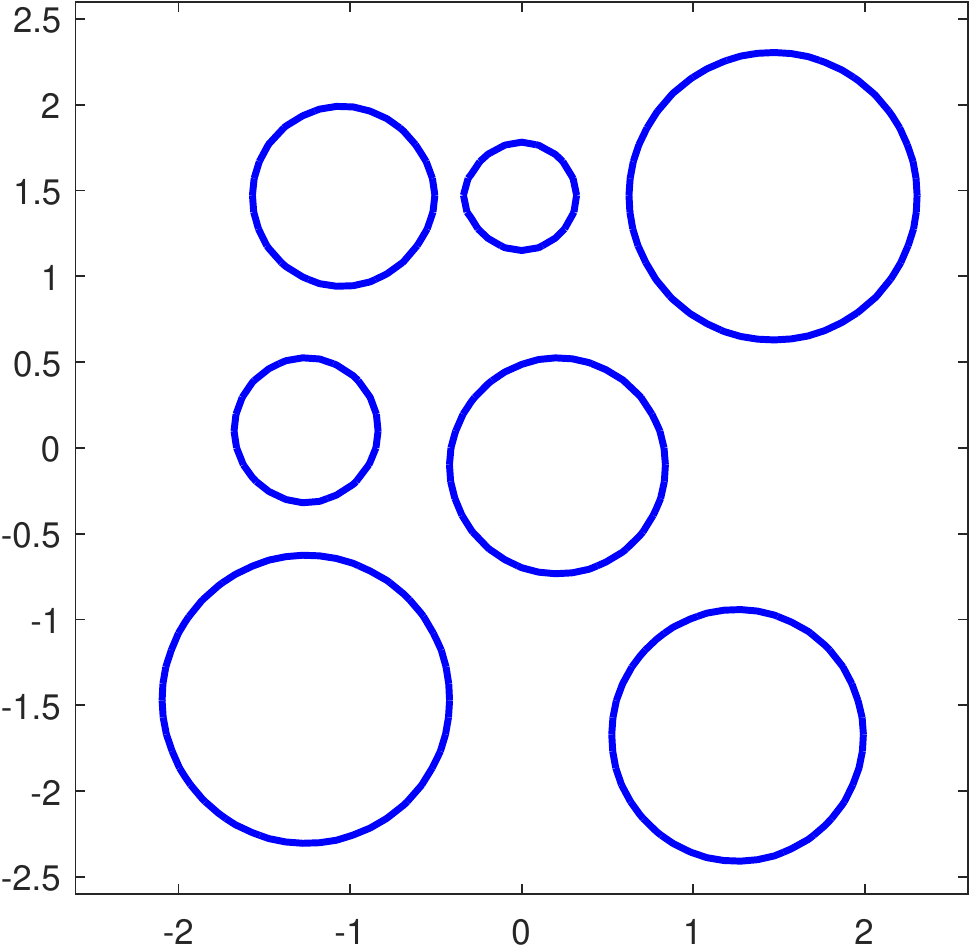}}
\subfigure[]{\includegraphics[width=1.8in]{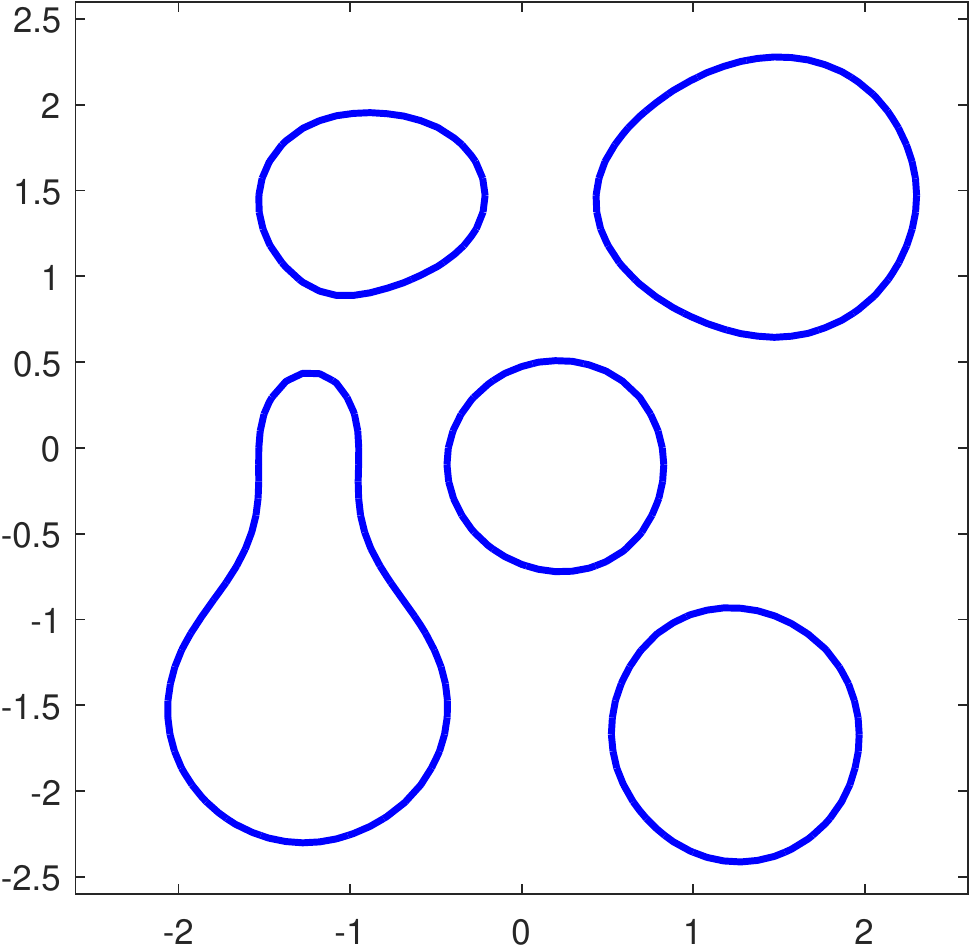}}
\subfigure[]{\includegraphics[width=1.8in]{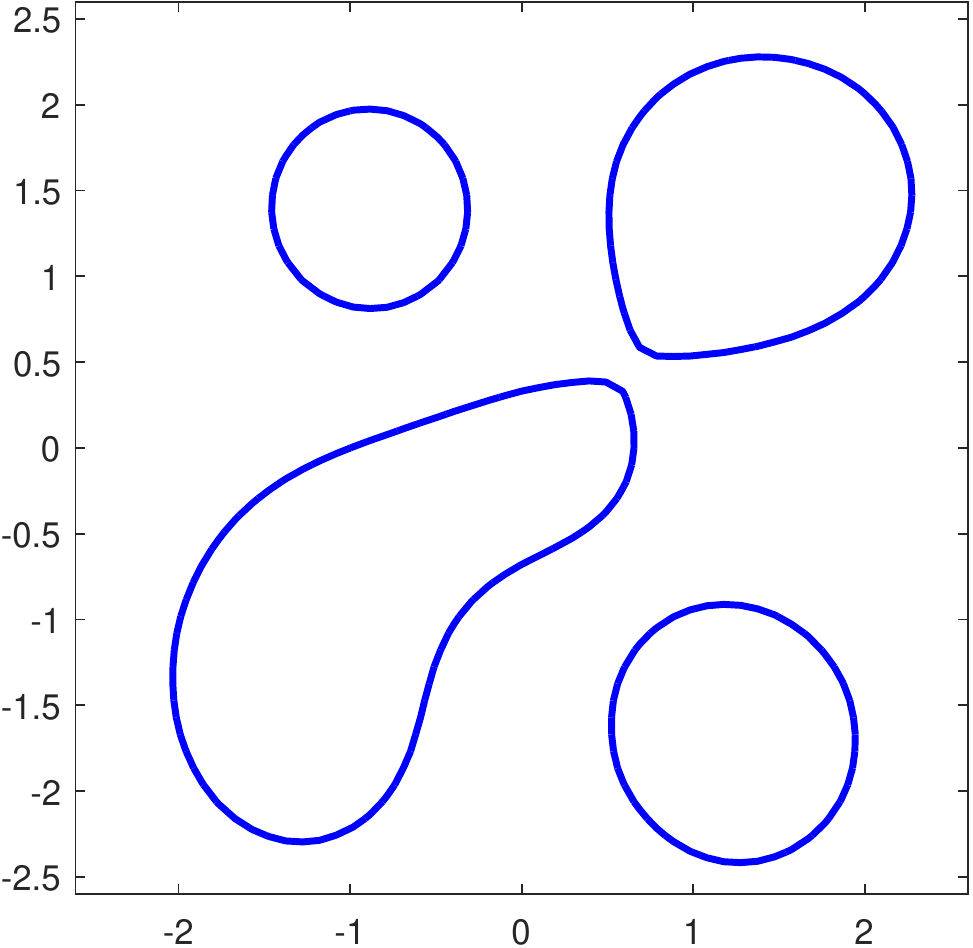}}
\subfigure[]{\includegraphics[width=1.8in]{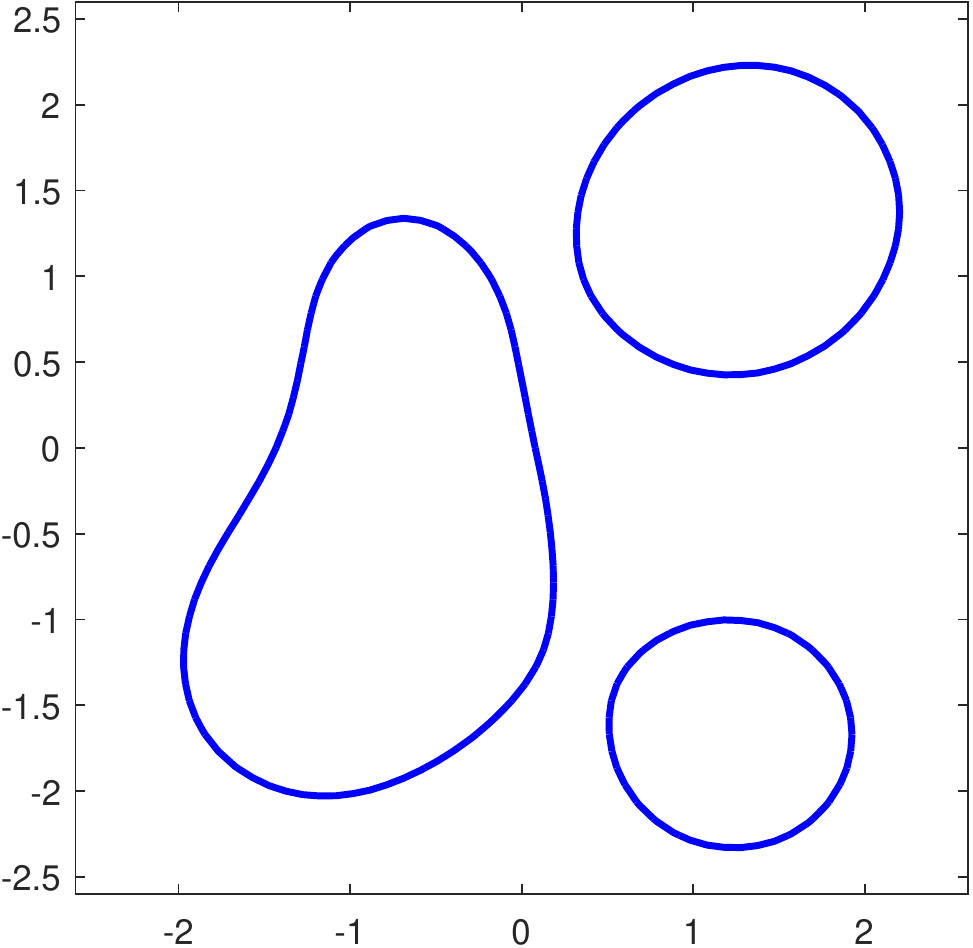}}
\subfigure[]{\includegraphics[width=1.8in]{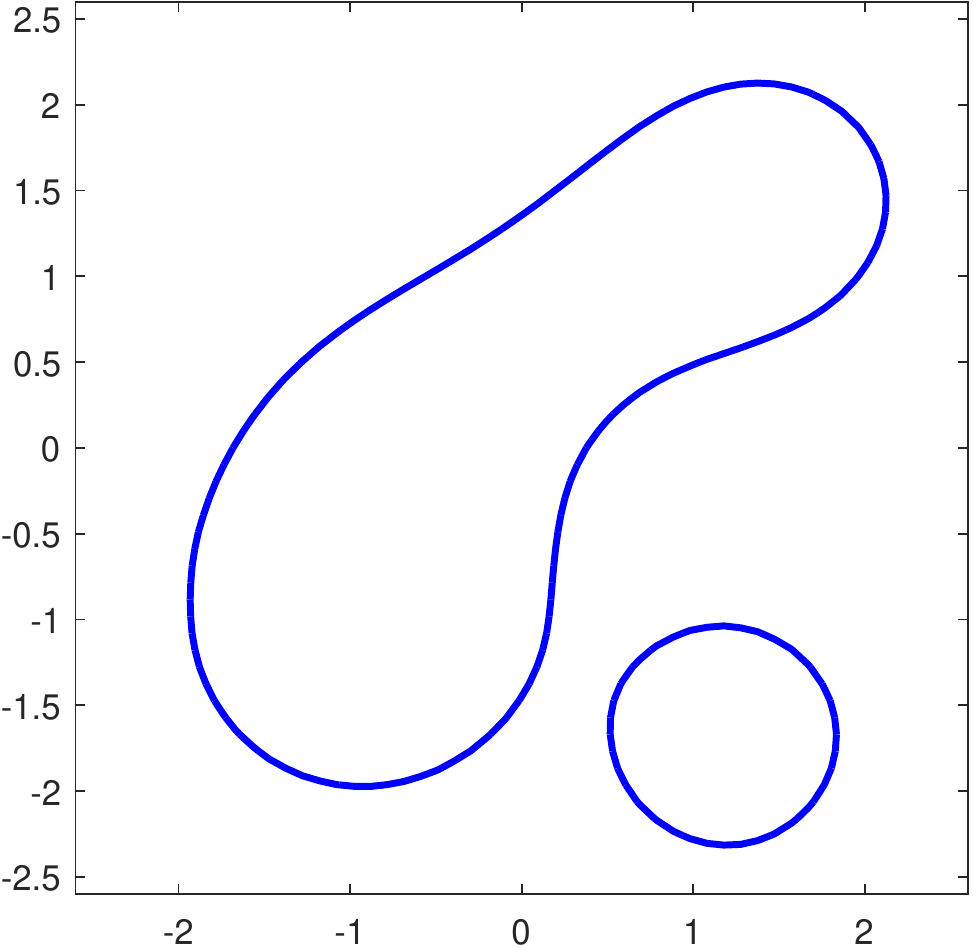}}
\subfigure[]{\includegraphics[width=1.8in]{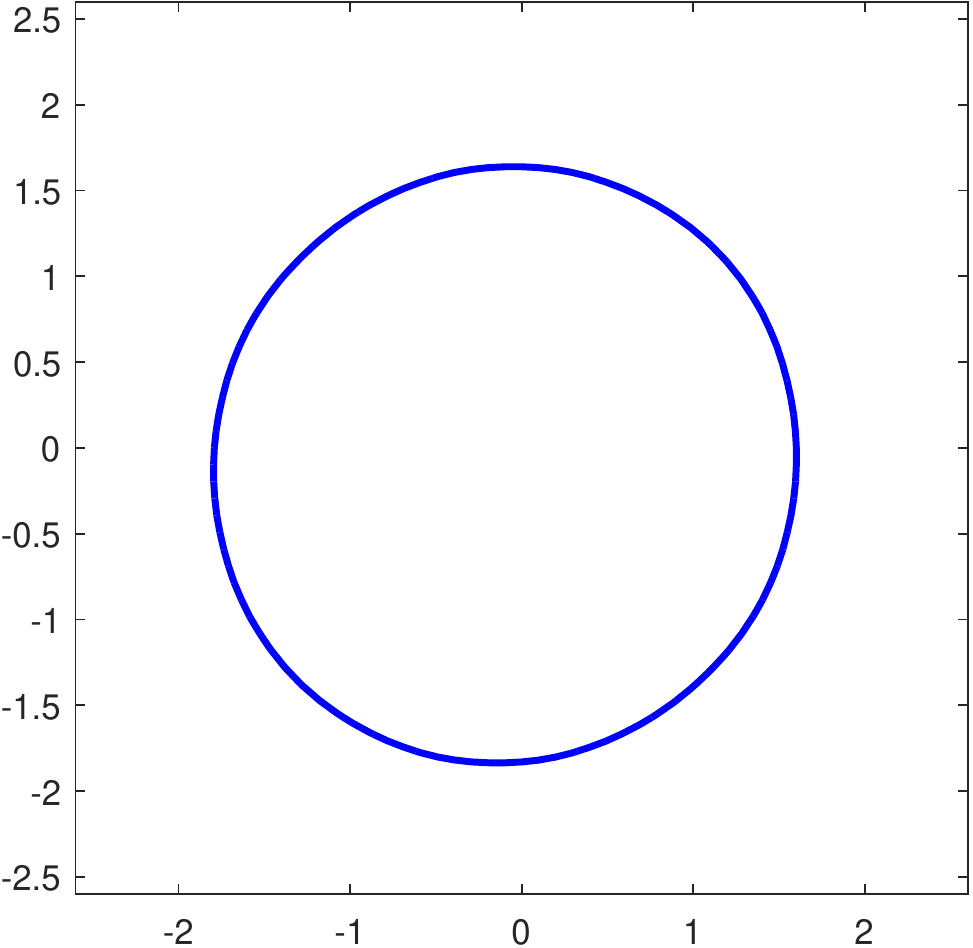}}
	\caption{Coalescence of seven prismatic loops by self-climb under their elastic interaction obtained by the phase field model.}\label{fig.7loops}
\end{figure}

Again, the topological changes during the coalescence process are handled automatically by the phase filed model without extra numerical treatment. Whereas if the front-tracking DDD method is used, the locations where two loops are merged have to be searched and a very small time step has to be used once such a location is identified, which will be very time-consuming for this simulation with frequent topological changes.

\subsection{Repelling of  prismatic loops with opposite directions}

\begin{figure}[htbp]
\centering
\includegraphics[width=2.5in]{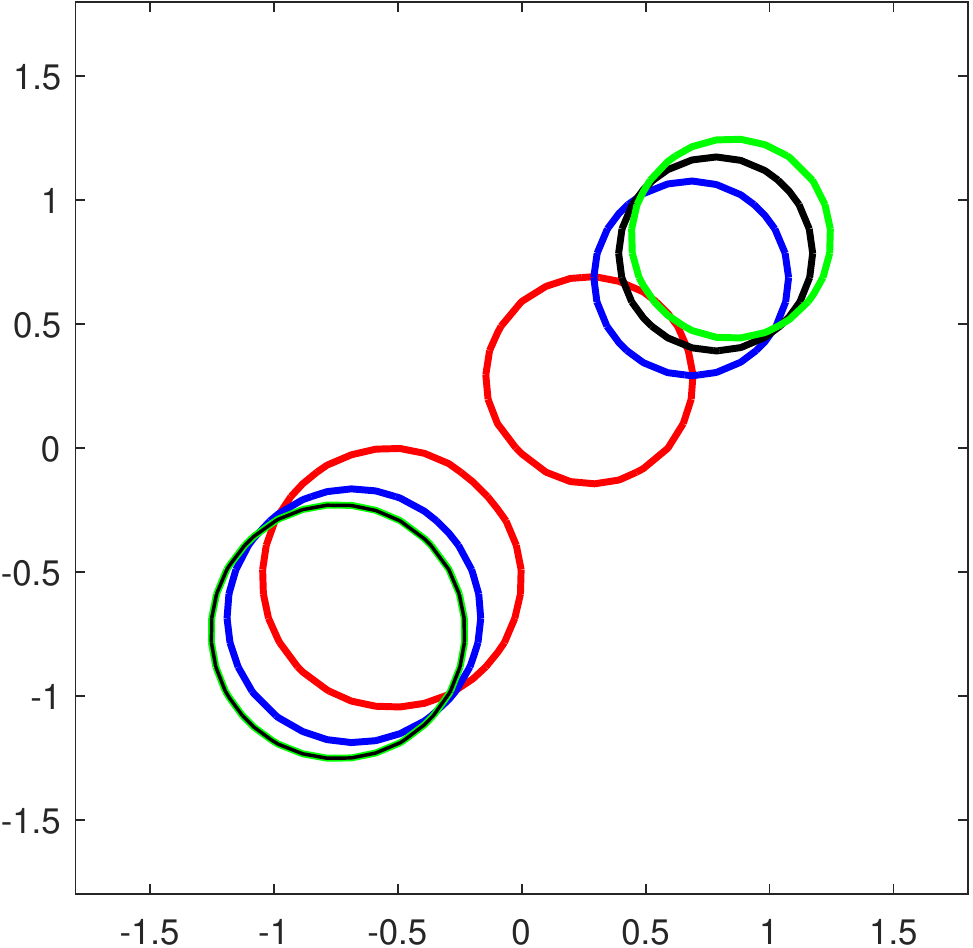}
	\caption{Repelling of two circular prismatic loops by self-climb under their elastic interaction obtained by our phase field model. The two red loops are the initial locations. Configurations of the two loops at different times during the evolution are shown by different colors.  }\label{fig.2loops}
\end{figure}

We simulate evolution of two prismatic loops with opposite directions under their elastic interaction, using the phase field model in Eqs.~\eqref{eqn.phasefield_multi1}--\eqref{eqn:mu_c2} with two order parameters $\phi_1$ and $\phi_2$. Initially,  one circular loop  has radius $20b$ with center $(13b, 13b)$, and is in the counterclockwise direction (vacancy loop) represented by $\phi_1$; the other circular loop  has radius $25b$ with center $(-25b, -25b)$, and is in the clockwise direction (interstitial loop) represented by $\phi_2$.
Evolution of the two prismatic loops is shown in  Fig.~\ref{fig.2loops}. It can be seen that the two loops are repelling each other by self-climb under their elastic interaction. Since the interaction force between the two loops decays as they move away from each other, the repelling motion of the two loops is fast initially and gradually slows down.


\begin{figure}[htbp]
\centering
\subfigure[]{\includegraphics[width=2.2in]{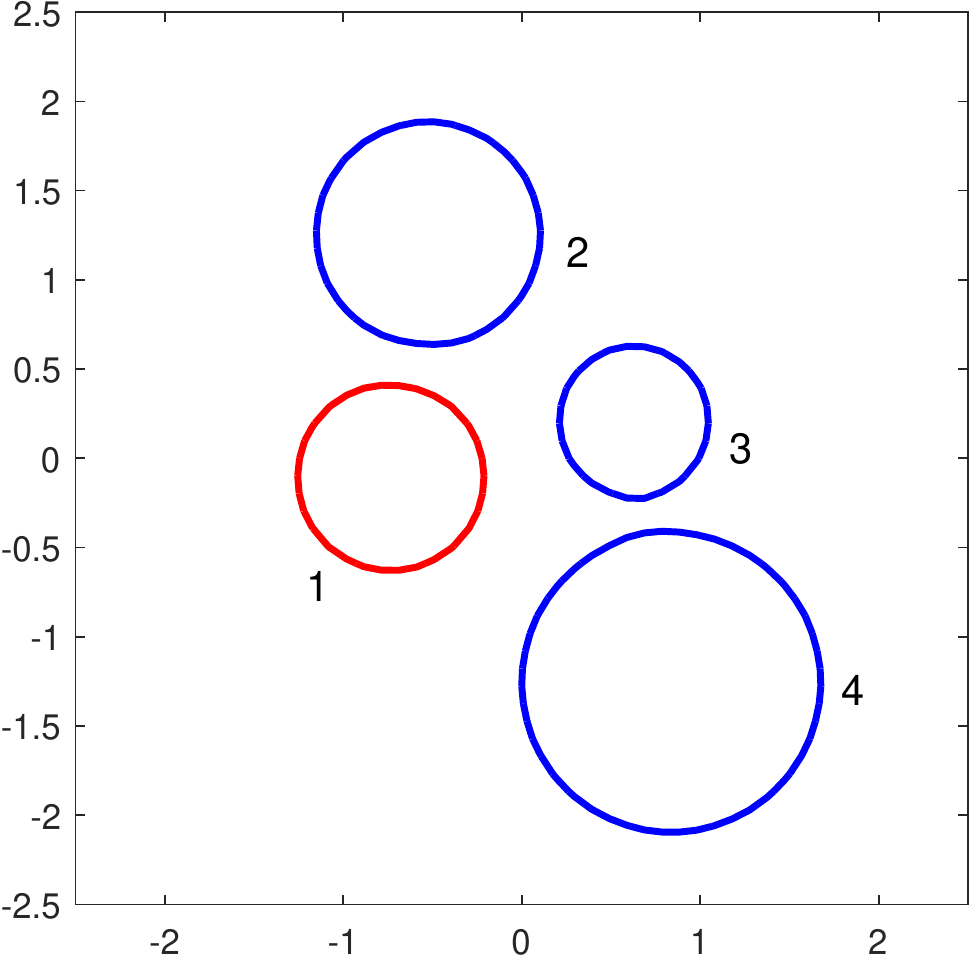}}
\subfigure[]{\includegraphics[width=2.2in]{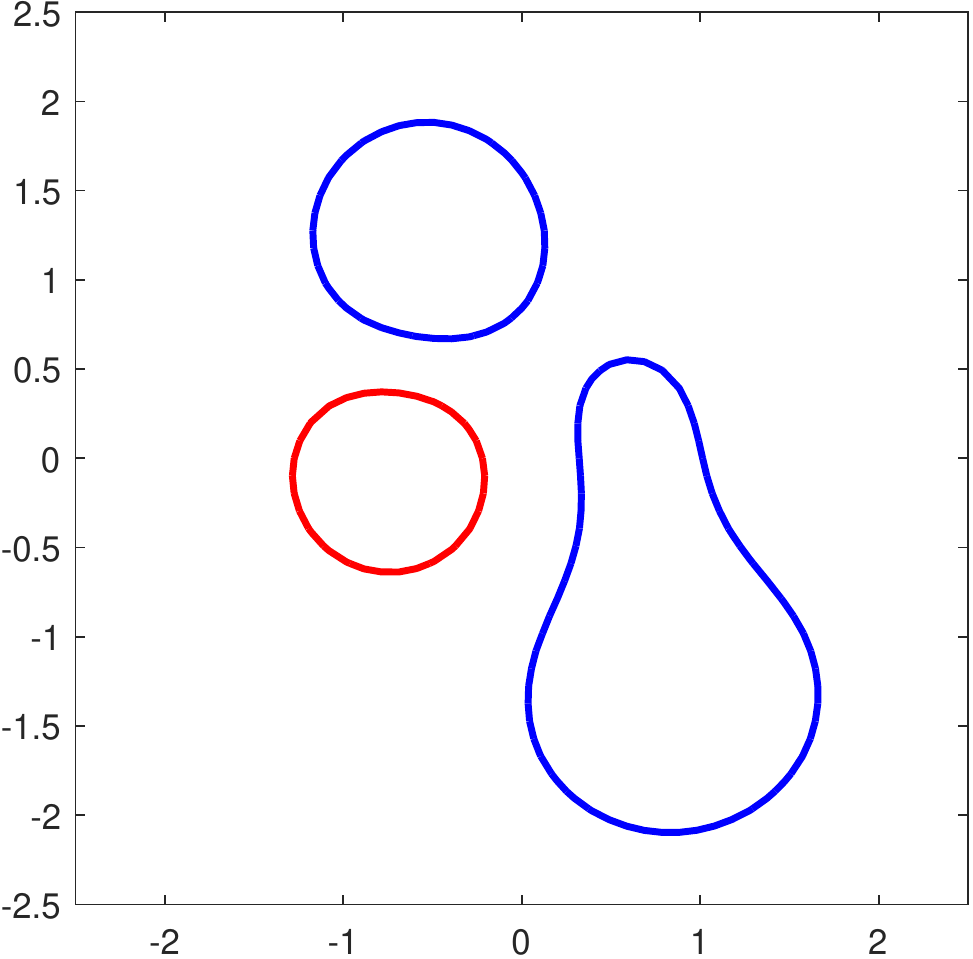}}
\subfigure[]{\includegraphics[width=2.2in]{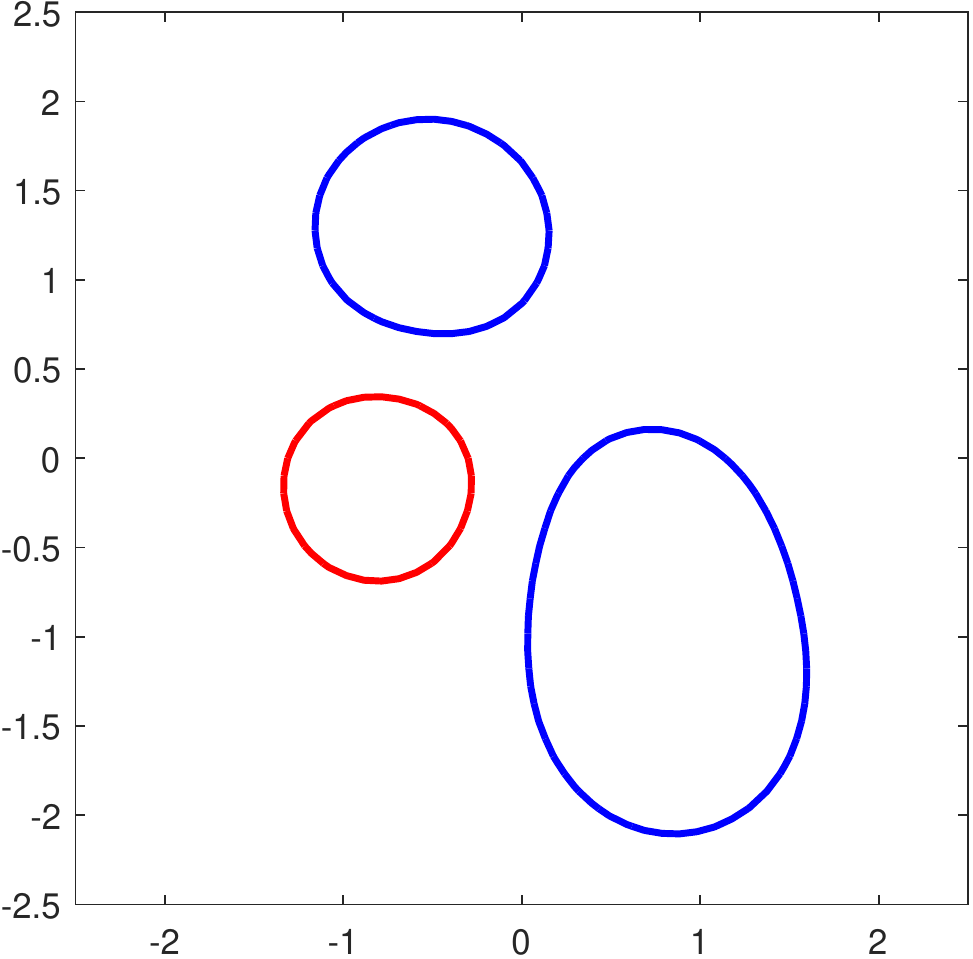}}
\subfigure[]{\includegraphics[width=2.2in]{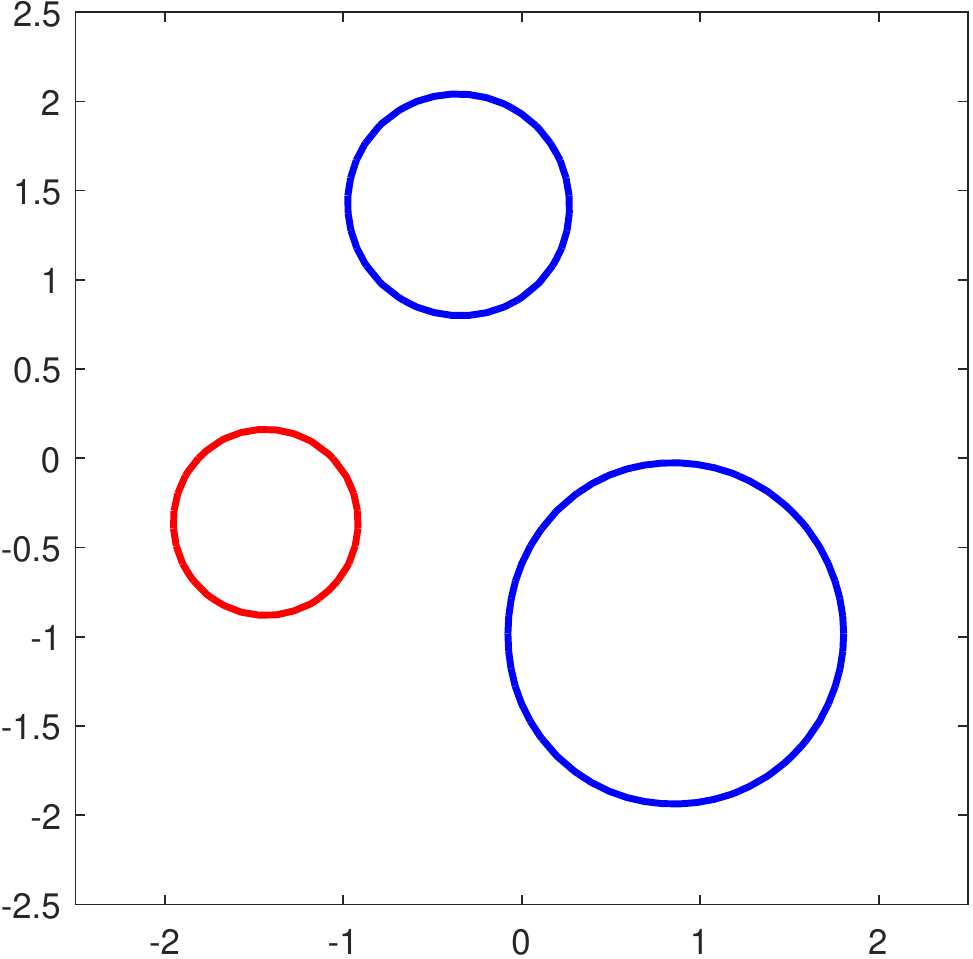}}
	\caption{Evolution of multiple prismatic loops with different directions by self-climb under their elastic interaction obtained by our phase field model. (a) Initial configuration: Loop 1 (in red) has the clockwise direction, while loops 2,3 and 4 (in blue) have the counterclockwise direction. (b)-(d) Snapshots during the evolution. }\label{fig.7loops}
\end{figure}

We also perform a simulation for the evolution of multiple prismatic loops with different directions; see Fig.~\ref{fig.7loops}. The initial configuration is shown in Fig.~\ref{fig.7loops}(a): Loop 1 (in red) has the clockwise direction, while loops 2,3 and 4 (in blue) have the counterclockwise direction.  During the evolution, loop 1 is repelled from the other three loops which have opposite direction. Loops 3 and 4 have the same direction and are attracted to each other and combined. Loop 2 is prevented from being attracted to loops 3 and 4 (or the one after combined) due to the repulsive effect from loop 1.

 \section{Conclusions and discussion}\label{sec:conclusions}

 In this paper, we have presented a new phase field model  for the self-climb of prismatic dislocation loops by vacancy pipe diffusion based on the  dislocation self-climb velocity formulation obtained in \cite{Niu2017}.  We focus on the two dimensional problem that the prismatic dislocation loops  lie and evolve by self-climb in a climb plane.  This phase field model is obtained by incorporating the climb force on dislocations into  the framework of the Cahn-Hilliard equation with degenerate mobility.
 Compared with the front-track DDD implementation method~\cite{Niu2019}, the developed phase field model has the advantage of being able to handle topological changes automatically and to avoid re-meshing during simulations.

 We have performed asymptotic analysis to show  that our phase field model yields the correct self-climb velocity  in the sharp interface limit.  In particular, we show that incorporation of the climb force in the framework of the classical Cahn-Hilliard equation  does not lead to significant change in the interface profile  (dislocation core profile here) of  the classical Cahn-Hilliard phase field model, and the framework of the Cahn-Hilliard equation gives a  curvature term that serves to correct the dislocation climb force calculated by the phase field formulation.

 We have validated our phase field model by numerical simulations and comparisons with the DDD simulation results obtained in  Ref.~\cite{Niu2019} and available  experimental observations.
  The simulations are performed to simulate the prismatic loop translation under the constant stress gradient, the evolution of the elliptic loop, the combination of several loops and the repelling of two loops with the different directions. The simulation results agree well with the DDD simulation results and experimental observations.
  \section*{Acknowledgments}
  The research is partly supported by National Natural Science Foundation of China under the
grant number 11801214. X. Yan's research is supported by a Research Excellence Grant from University of Connecticut.
 \bibliographystyle{plainnat}

\bibliography{mybib}

\end{document}